\documentclass[aps,prd,twocolumn,showpacs,notitlepage,eqsecnum,nofootinbib,
superscriptaddress]{revtex4-1}

\usepackage[centertags]{amsmath}
\usepackage{amssymb}
\usepackage{latexsym}
\usepackage{enumerate}
\usepackage{graphicx}
\usepackage{mathrsfs}
\usepackage{hyperref}
\usepackage{stmaryrd}
\usepackage{import}
\usepackage{tensor}
\usepackage{color}
\usepackage{bm}
\usepackage{multirow}
\usepackage{breakurl}
\usepackage{float}
\usepackage{placeins}

\newcommand{\bi}{\begin{itemize}}
\newcommand{\ei}{\end{itemize}}
\newcommand{\be}{\begin{equation}}
\newcommand{\ee}{\end{equation}}

\renewcommand{\l}{\left(}
\renewcommand{\r}{\right)}
\renewcommand{\a}{\alpha}

\renewcommand{\d}{\delta}

\newcommand{\la}{\lambda}
\renewcommand{\O}{\Omega}
\renewcommand{\o}{\omega}
\renewcommand{\th}{\theta}

\newcommand{\q}{\quad}

\newcommand{\vp}{\varphi}

\newcommand{\pa}{\partial}

\allowdisplaybreaks

\begin{document}

\title{Eccentric-orbit EMRI gravitational wave energy fluxes to 7PN order} 

\author{Erik Forseth}
\author{Charles R. Evans}
\affiliation{Department of Physics and Astronomy, University of North 
Carolina, Chapel Hill, North Carolina 27599, USA}
\author{Seth Hopper}
\affiliation{School of Mathematics and Statistics and Complex \& Adaptive 
Systems Laboratory, University College Dublin, Belfield, Dublin 4, Ireland}
\affiliation{CENTRA, Departamento de F\'{i}sica, Instituto 
Superior T\'{e}cnico – IST, Avenida Rovisco Pais 1, 1049, Lisboa, Portugal}

\begin{abstract}
We present new results through 7PN order on the energy flux from eccentric 
extreme-mass-ratio binaries.  The black hole perturbation calculations are 
made at very high accuracy (200 decimal places) using a \emph{Mathematica} 
code based on the Mano-Suzuki-Takasugi (MST) analytic function expansion 
formalism.  All published coefficients in the expansion through 3PN order 
at lowest order in the mass ratio are confirmed and new analytic and numeric 
terms are found to high order in powers of $e^2$ at post-Newtonian orders 
between 3.5PN and 7PN.  We also show original work in finding (nearly) 
arbitrarily accurate expansions for 
hereditary terms at 1.5PN, 2.5PN, and 3PN orders.  An asymptotic analysis is 
developed that guides an understanding of eccentricity singular factors, 
which diverge at unit eccentricity and which appear at each PN order.  We 
fit to a model at each PN order that includes these eccentricity singular 
factors, which allows the flux to be accurately determined out to $e \to 1$.
\end{abstract}


\maketitle

\section{Introduction}
\label{sec:intro}

Merging compact binaries have long been thought to be promising sources of 
gravitational waves that might be detectable in ground-based (Advanced LIGO, 
Advanced VIRGO, KAGRA, etc) \cite{LIGO,VIRGO,KAGRA} or space-based (eLISA) 
\cite{eLISA} experiments.  With the first observation of a binary black 
hole merger (GW150914) by Advanced LIGO \cite{AbboETC16a}, the era of 
gravitational wave astronomy has arrived.  This first observation emphasizes 
what was long understood--that detection of weak signals and physical parameter 
estimation will be aided by accurate theoretical predictions.  Both the native 
theoretical interest and the need to support detection efforts combine to 
motivate research in three complementary approaches \cite{Leti14} for 
computing merging binaries: numerical relativity \cite{BaumShap10,LehnPret14}, 
post-Newtonian (PN) theory \cite{Will11,Blan14}, and gravitational self-force 
(GSF)/black hole perturbation (BHP) calculations 
\cite{DrasHugh05,Bara09,PoisPounVega11, Thor11,Leti14}.  The effective-one-body 
(EOB) formalism then provides a synthesis, drawing calibration of its 
parameters from all three \cite{BuonDamo99,BuonETC09,Damo10,HindETC13,
Damo13,TaraETC14}.  

\begin{figure}
\includegraphics[scale=0.95]{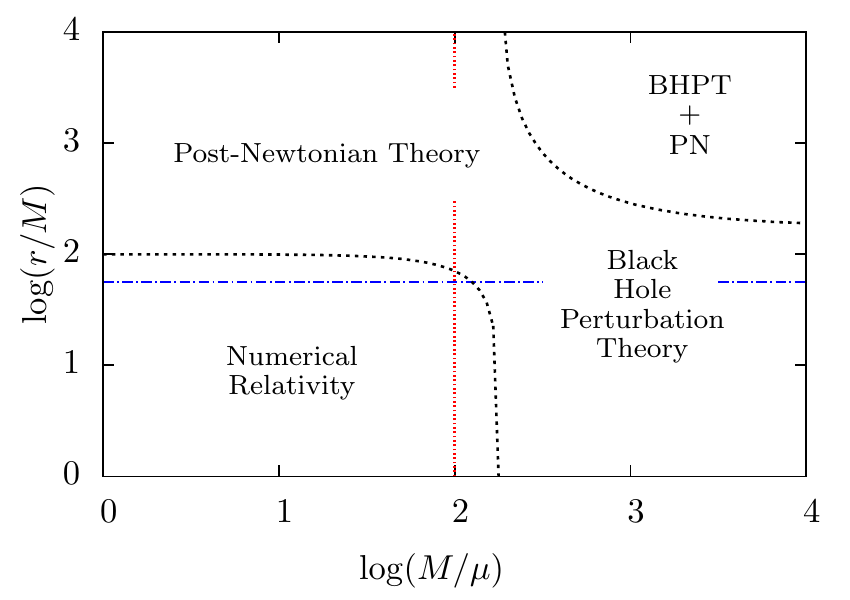}
\caption{Regions of binary parameter space in which different formalisms 
apply.  Post-Newtonian (PN) approximation applies best to binaries with wide 
orbital separation (or equivalently low frequency).  Black hole perturbation 
(BHP) theory is relevant for binaries with small mass ratio $\mu/M$.  
Numerical relativity (NR) works best for close binaries with comparable 
masses.  This paper makes comparisons between PN and BHP results in their 
region of mutual overlap. 
\label{fig:regimes}} 
\end{figure}

In the past seven years numerous comparisons \cite{Detw08,SagoBaraDetw08,
BaraSago09,BlanETC09,BlanETC10,Fuji12,ShahFrieWhit14,Shah14,
JohnMcDaShahWhit15,AkcaETC15} have been made in the overlap region 
(Fig.~\ref{fig:regimes}) between GSF/BHP theory and PN theory.  PN 
theory is accurate for orbits with wide separations (or low frequencies) but 
arbitrary component masses, $m_1$ and $m_2$.  The GSF/BHP approach assumes a 
small mass ratio $q = m_1/m_2 \ll 1$ (notation typically being $m_1 = \mu$ 
with black hole mass $m_2 = M$).  While requiring small $q$, GSF/BHP 
theory has no restriction on orbital separation or field strength.  Early BHP 
calculations focused on comparing energy fluxes; see for example 
\cite{Pois93,CutlETC93,TagoSasa94,TagoNaka94} for waves radiated to infinity 
from circular orbits and \cite{PoisSasa95} for flux absorbed at the black hole 
horizon.  Early calculations of losses from eccentric orbits were made by 
\cite{TanaETC93,AposETC93,CutlKennPois94,Tago95}.  More recently, starting 
with Detweiler \cite{Detw08}, it became possible with GSF theory to compare 
\emph{conservative} gauge-invariant quantities \cite{SagoBaraDetw08,
BaraSago09,BlanETC09,BlanETC10,ShahFrieWhit14,DolaETC14b,JohnMcDaShahWhit15,
BiniDamoGera15,AkcaETC15,HoppKavaOtte15}.  With the advent of extreme 
high-accuracy GSF calculations \cite{Fuji12,ShahFrieWhit14} focus also 
returned to calculating dissipative effects (fluxes), this time to 
extraordinarily high PN order \cite{Fuji12,Shah14} for circular orbits.  This 
paper concerns itself with making similar extraordinarily accurate (200 
digits) calculations to probe high PN order energy flux from eccentric orbits.

The interest in eccentric orbits stems from astrophysical considerations 
\cite{AmarETC07,AmarETC14} that indicate extreme-mass-ratio inspirals (EMRIs) 
should be born with high eccentricities.  Other work \cite{HopmAlex05} 
suggests EMRIs will have a distribution peaked about $e = 0.7$ as they enter 
the eLISA passband.  Less extreme (intermediate) mass ratio inspirals (IMRIs) 
may also exist \cite{MillColb04} and might appear as detections in Advanced 
LIGO \cite{BrowETC07,AmarETC07}.  Whether they exist, and have significant 
eccentricities, is an issue for observations to settle.  The PN expansion for 
eccentric orbits is known through 3PN relative order \cite{ArunETC08a,
ArunETC08b,ArunETC09a,Blan14}.  The present paper confirms the accuracy of 
that expansion for the energy flux and determines PN eccentricity-dependent 
coefficients all the way through 7PN order for multiple orders in an expansion 
in $e^2$.  The model is improved by developing an understanding of what 
eccentricity singular functions to factor out at each PN order.  In so doing, 
we are able to obtain better convergence and the ability to compute the 
flux even as $e \rightarrow 1$.  The review by Sasaki and Tagoshi 
\cite{SasaTago03} summarized earlier work on fluxes from slightly eccentric 
orbits (through $e^2$) and more recently results have been obtained 
\cite{SagoFuji15} on fluxes to $e^6$ for 3.5PN and 4PN order.

Our work makes use of the analytic function expansion formalism developed 
by Mano, Suzuki, and Takasugi (MST) \cite{ManoSuzuTaka96a,ManoSuzuTaka96b} 
with a code written in \emph{Mathematica} (to take advantage of arbitrary 
precision functions).  The MST formalism expands solutions to the Teukolsky 
equation in infinite series of hypergeometric functions.  We convert from 
solutions to the Teukolsky equation to solutions of the Regge-Wheeler-Zerilli 
equations and use techniques found in \cite{HoppEvan10,HoppETC15}.  Our use 
of MST is similar to that found in Shah, Friedman, and Whiting 
\cite{ShahFrieWhit14}, who studied conservative effects, and Shah 
\cite{Shah14}, who examined fluxes for circular equatorial orbits on Kerr.  

This paper is organized as follows.  Those readers interested primarily in new 
PN results will find them in Secs.~\ref{sec:preparePN}, \ref{sec:confirmPN}, 
and \ref{sec:newPN}.  Sec.~\ref{sec:preparePN} contains original work in 
calculating the 1.5PN, 2.5PN, and 3PN hereditary terms to exceedingly high 
order in powers of the eccentricity to facilitate comparisons with 
perturbation theory.  It includes a subsection, Sec.~\ref{sec:asymptotic}, 
that uses an asymptotic analysis to guide an understanding of different 
eccentricity singular factors that appear in the flux at all PN orders.  
In Sec.~\ref{sec:confirmPN} we verify all previously known PN coefficients 
(i.e., those through 3PN relative order) in the energy flux from eccentric 
binaries at lowest order in the mass ratio.  Sec.~\ref{sec:newPN} and 
App.~\ref{sec:numericEnh} present our new findings on PN coefficients in the 
energy flux from eccentric orbits between 3.5PN and 7PN order.  For those 
interested in the method, Sec.~\ref{sec:homog} reviews the MST formalism for 
analytic function expansions of homogeneous solutions, and describes the 
conversion from Teukolsky modes to normalized Regge-Wheeler-Zerilli modes.  
Section \ref{sec:InhomogSol} outlines the now-standard procedure of solving 
the RWZ source problem with extended homogeneous solutions, though now with 
the added technique of spectral source integration \cite{HoppETC15}.  Some 
details on our numerical procedure, which allows calculations to better than 
200 decimal places of accuracy, are given in Sec.~\ref{sec:CodeDetails}.  Our 
conclusions are drawn in Sec.~\ref{sec:conclusions}.

Throughout this paper we set $c = G = 1$, and use metric signature $(-+++)$ 
and sign conventions of Misner, Thorne, and Wheeler \cite{MisnThorWhee73}.  
Our notation for the RWZ formalism is that of \cite{HoppEvan10}, which derives 
from earlier work of Martel and Poisson \cite{MartPois05}.  

\section{Analytic expansions for homogeneous solutions}
\label{sec:homog}

This section briefly outlines the MST formalism \cite{ManoSuzuTaka96b} (see 
the detailed review by Sasaki and Tagoshi \cite{SasaTago03}) and describes 
our conversion to analytic expansions for normalized RWZ modes.

\subsection{The Teukolsky formalism}

The MST approach provides analytic function expansions for general 
perturbations of a Kerr black hole.  With other future uses in mind, elements 
of our code are based on the general MST expansion.  However, the present 
application is focused solely on eccentric motion in a Schwarzschild 
background and thus in our discussion below we simply adopt the $a = 0$ limit 
on black hole spin from the outset.  The MST method describes gravitational 
perturbations in the Teukolsky formalism \cite{Teuk73} using the 
Newman-Penrose scalar 
$\psi_4 = - C_{\alpha\beta\gamma\delta} n^{\alpha} \bar{m}^{\beta} 
n^{\gamma} \bar{m}^{\delta}$ \cite{NewmPenr62,NewmPenr63}.  Here  
$C_{\alpha\beta\gamma\delta}$ is the Weyl tensor, and its projection is made 
on elements of the Kinnersley null tetrad (see \cite{Kinn69,Teuk73} for its 
components). 

In our application the line element is 
\be
ds^2 = -f dt^2 + f^{-1} dr^2
+ r^2 \left( d\theta^2 + \sin^2\theta \, d\varphi^2 \right) ,
\ee
as written in Schwarzschild coordinates, with $f(r) = 1 - 2M/r$.  The 
Teukolsky equation \cite{Teuk73} with spin-weight $s = -2$ is satisfied 
(when $a=0$) by $r^4 \psi_4$, with $\psi_4$ separated into Fourier-harmonic 
modes by
\be
\psi_4 = r^{-4} \sum_{lm} \, \int \, d\o \, e^{-i\o t} 
\, R_{lm\o}(r) \, {}_{-2}Y_{lm}(\th, \vp) .
\ee
Here ${}_{s}Y_{lm}$ are spin-weighted spherical harmonics.  The Teukolsky 
equation for $R_{lm\o}$ reduces in our case to the Bardeen-Press equation 
\cite{BardPres73,CutlKennPois94}, which away from the source has the 
homogeneous form
\be
\label{eqn:radial}
\left[ r^2 f \frac{d^2}{dr^2} - 2(r - M) \frac{d}{dr} + U_{l\omega}(r)\right]
\, R_{lm\omega}(r) = 0 ,
\ee
with potential
\be
U_{l\omega}(r) = \frac{1}{f}\left[\omega^2 r^2 - 4i\omega (r - 3M)\right] 
- (l-1)(l+2) .
\ee

Two independent homogeneous solutions are of interest, which have, 
respectively, causal behavior at the horizon, $R^{\rm in}_{lm\omega}$, and at 
infinity, $R^{\rm up}_{lm\omega}$,
\begin{align}
\label{eqn:Rin}
\hspace{-1.95ex}
R^{\rm in}_{lm\omega} &= 
\begin{cases}
B^{\rm trans}_{lm\omega} r^2 f \, e^{-i \o r_*} \, &
r \rightarrow 2M
\\
B^{\rm ref}_{lm\omega} r^3 \, e^{i\omega r_*} + 
\frac{B^{\rm in}_{lm\omega}}{r} \, e^{-i\omega r_*} \, &
r \rightarrow +\infty ,
\end{cases}
\\
\label{eqn:Rup}
\hspace{-1.95ex}
R^{\rm up}_{lm\omega} &= 
\begin{cases}
C^{\rm up}_{lm\omega} \, e^{i\o r_*} + 
C^{\rm ref}_{lm\omega} r^2 f \, e^{-i\o r_*} \, &
r \rightarrow 2M
\\
C^{\rm trans}_{lm\omega} r^3 \, e^{i\omega r_*} \, &
r \rightarrow +\infty ,
\end{cases}
\end{align}
where $B$ and $C$ are used for incident, reflected, and transmitted 
amplitudes.  Here $r_*$ is the usual Schwarzschild tortoise coordinate
$r_* = r + 2M \log (r/2M - 1 )$.  

\subsection{MST analytic function expansions for $R_{lm\o}$}
\label{sec:MST}

The MST formalism makes separate analytic function expansions for the 
solutions near the horizon and near infinity.  We begin with the near-horizon 
solution.

\subsubsection{Near-horizon (inner) expansion}
\label{sec:innerMST}

After factoring out terms that arise from the existence of singular points, 
$R^{\rm in}_{lm\omega}$ is represented by an infinite series in hypergeometric 
functions
\begin{align}
\label{eqn:Down1}
R_{lm\omega}^{\text{in}} &= 
e^{i\epsilon x}(-x)^{2-i\epsilon}
p_\text{in}^{\nu}(x) ,
\\
\label{eqn:DownSeries}
p_\text{in}^\nu(x) &= \sum_{n=-\infty}^{\infty} a_n p_{n+\nu}(x) ,
\end{align}
where $\epsilon = 2M\omega$ and $x = 1 - r/2M$.  The functions $p_{n+\nu}(x)$ 
are an alternate notation for the hypergeometric functions 
${}_2F_1(a,b;c;x)$, with the arguments in this case being
\be
\label{eqn:DownPDef}
p_{n+\nu}(x) = {}_2F_1(n+\nu+1-i\epsilon,-n-\nu-i\epsilon;3-2i\epsilon;x) .
\ee
The parameter $\nu$ is freely specifiable and referred to as the 
\emph{renormalized angular momentum,} a generalization of $l$ to non-integer 
(and sometimes complex) values.  

The series coefficients $a_n$ satisfy a three-term recurrence relation
\be
\label{eqn:recurrence}
\alpha_n^\nu a_{n+1} + \beta_n^\nu a_n + \gamma_n^\nu a_{n-1} = 0 ,
\ee
where $\alpha_n^\nu$, $\beta_n^\nu$, and $\gamma_n^\nu$ depend on $\nu$, $l$, 
$m$, and $\epsilon$ (see App.~\ref{sec:solveNu} and 
Refs.~\cite{ManoSuzuTaka96b} and \cite{SasaTago03} for details).  The 
recurrence relation has two linearly-independent solutions, $a_n^{(1)}$ and 
$a_n^{(2)}$.  Other pairs of solutions, say $a_n^{(1')}$ and $a_n^{(2')}$, 
can be obtained by linear transformation.  Given the asymptotic form of 
$\alpha_n^\nu$, $\beta_n^\nu$, and $\gamma_n^\nu$, it is possible to find 
pairs of solutions such that 
$\lim_{n\rightarrow +\infty} a_n^{(1)}/a_{n}^{(2)} = 0$ and 
$\lim_{n\rightarrow -\infty} a_n^{(1')}/a_{n}^{(2')} = 0$.  The two 
sequences $a_n^{(1)}$ and $a_n^{(1')}$ are called \emph{minimal} solutions 
(while $a_n^{(2)}$ and $a_n^{(2')}$ are \emph{dominant} solutions), but in 
general the two sequences will not coincide.  This is where the free 
parameter $\nu$ comes in.  It turns out possible to choose $\nu$ such that 
a unique minimal solution emerges (up to a multiplicative constant), with 
$a_n(\nu)$ uniformly valid for $-\infty < n < \infty$ and with the series 
converging.  The procedure for finding $\nu$, which depends on frequency, 
and then finding $a_n(\nu)$, involves iteratively solving for the root of 
an equation that contains 
continued fractions and resolving continued fraction equations.  We give
details in App.~\ref{sec:solveNu}, but refer the reader to \cite{SasaTago03} 
for a complete discussion.  The expansion for $R_{lm\o}^{\rm in}$ converges 
everywhere except $r=\infty$.  For the behavior there we need a separate 
expansion.

\subsubsection{Near-infinity (outer) expansion}
\label{sec:outerMST}

After again factoring out terms associated with singular points, an infinite 
expansion can be written \cite{ManoSuzuTaka96b,SasaTago03,Leav86} for the 
outer solution $R^{\rm up}_{lm\o}$ with outgoing wave dependence,
\begin{align}
\label{eqn:RMinus}
R^{\rm up}_{lm\o} & = 2^\nu e^{-\pi\epsilon}e^{-i\pi(\nu-1)}e^{iz}
z^{\nu+i\epsilon}(z-\epsilon)^{2-i\epsilon} \\
&\times\sum_{n=-\infty}^\infty i^n
\frac{(\nu-1-i\epsilon)_n}{(\nu+3+i\epsilon)_n}
b_n (2z)^n \notag \\
&\qquad\qquad\times
\Psi(n+\nu-1-i\epsilon,2n+2\nu+2;-2iz) . \notag
\end{align}
Here $z = \o r = \epsilon (1 - x)$ is another dimensionless variable, 
$(\zeta)_n = \Gamma(\zeta+n)/\Gamma(\zeta)$ is the (rising) Pochhammer 
symbol, and $\Psi(a,c;x)$ are irregular confluent hypergeometric functions.
The free parameter $\nu$ has been introduced again as well.   The limiting 
behavior $ \lim_{|x|\rightarrow\infty}\Psi(a,c;x)\rightarrow x^{-a}$ 
guarantees the proper asymptotic dependence
$ R^{\rm up}_{lm\o} = C^{\rm trans}_{lm\o} (z/\o)^{3} \, 
e^{i(z+\epsilon\log{z})}$.

Substituting the expansion in \eqref{eqn:radial} produces a three-term 
recurrence relation for $b_n$.  Remarkably, because of the Pochhammer 
symbol factors that were introduced in \eqref{eqn:RMinus}, the recurrence 
relation for $b_n$ is identical to the previous one \eqref{eqn:recurrence} 
for the inner solution.  Thus the same value for the renormalized angular 
momentum $\nu$ provides a uniform minimal solution for $b_n$, which can 
be identified with $a_n$ up to an arbitrary choice of normalization.

\subsubsection{Recurrence relations for homogeneous solutions}

Both the ordinary hypergeometric functions ${}_2F_1(a,b;c;z)$ and the
irregular confluent hypergeometric functions $\Psi(a,b;z)$ admit three term 
recurrence relations, which can be used to speed the construction of 
solutions \cite{Shah14b}.  The hypergeometric functions 
$p_{n+\nu}$ in the inner solution \eqref{eqn:DownSeries} satisfy
\begin{align}
\label{eqn:DowngoingRecurrence}
&p_{n+\nu} =  -\frac{2n+2\nu-1}{(n+\nu-1)(2+n+\nu-i\epsilon)(n+\nu-i\epsilon)}
\notag\\
&\hspace{3ex}\times\left[(n+\nu)(n+\nu-1)(2x-1)
+(2i+\epsilon)\epsilon\right]p_{n+\nu-1}\notag\\
&\hspace{3ex}-\frac{(n+\nu)(n+\nu+i\epsilon-3)(n+\nu+i\epsilon-1)}{(n+\nu-1)
(2+n+\nu-i\epsilon)(n+\nu-i\epsilon)} p_{n+\nu-2}.\notag \\
\end{align}
Defining by analogy with Eqn.~\eqref{eqn:DownPDef} 
\be
q_{n+\nu} \equiv \Psi(n+\nu-i\epsilon -1,2n+2\nu+2;-2iz) ,
\ee
the irregular confluent hypergeometric functions satisfy
\begin{align}
\label{eqn:OutgoingRecurrence}
& q_{n+\nu} = \frac{(2n+2\nu-1)}{(n+\nu-1)(n+\nu-i\epsilon - 2)z^2} \notag \\
&\hspace{3ex} \times\left[2n^2+2\nu(\nu-1)+n(4\nu-2)-(2+i\epsilon)z\right] 
q_{n+\nu-1} \notag \\
&\hspace{3ex}
+\frac{(n+\nu)(1+n+\nu+i\epsilon)}{(n+\nu-1)(n+\nu-i\epsilon-2)z^2}
q_{n+\nu-2}.
\end{align}

\subsection{Mapping to RWZ master functions}

In this work we map the analytic function expansions of $R_{lm\o}$ to ones 
for the RWZ master functions.  The reason stems from having pre-existing 
coding infrastructure for solving RWZ problems \cite{HoppEvan10} and the 
ease in reading off gravitational wave fluxes.  The Detweiler-Chandrasekhar 
transformation \cite{Chan75,ChanDetw75,Chan83} maps $R_{lm\o}$ to a solution 
$X^{\rm RW}_{lm\omega}$ of the Regge-Wheeler equation via
\be
\label{eqn:RWtrans1}
X^{\rm RW}_{lm\omega} = r^3
\left(\frac{d}{dr}-\frac{i\omega}{f}\right)
\left(\frac{d}{dr}-\frac{i\omega}{f}\right)
\frac{R_{lm\omega}}{r^2} .
\ee
For odd parity ($l + m =$ odd) this completes the transformation.  For even 
parity, we make a second transformation \cite{Bern07} to map through to a 
solution $X_{lm\omega}^Z$ of the Zerilli equation 
\begin{align}
\label{eqn:RWtrans2}
X_{lm\omega}^{Z,\pm} &= 
\frac{1}{\la (\la + 1) \pm 3 i \o M} 
\Bigg\{
3 M f\frac{dX_{lm\o}^{\rm{RW},\pm}}{dr} \\
& \hspace{7ex}+
\left[\la(\la+1)+\frac{9 M^2 f}{r(\la r+3M)}\right] 
X_{lm\o}^{\rm{RW},\pm}
 \Bigg\} . \notag
\end{align}
Here $\la=(l-1)(l+2)/2$.  We have introduced above the $\pm$ notation to 
distinguish outer ($+$) and inner ($-$) solutions--a notation that will be 
used further in Sec.~\ref{sec:TDmasterEq}.  [When unambiguous we often use 
$X_{lm\o}$ to indicate either the RW function (with $l+m=$ odd) or Zerilli 
function (with $l+m=$ even).]  The RWZ functions satisfy the 
homogeneous form of \eqref{eqn:masterInhomogFD} below with their respective 
parity-dependent potentials $V_l$.

The normalization of $R_{lm\o}$ in the MST formalism is set by adopting some 
starting value, say $a_0 = 1$, in solving the recurrence relation for $a_n$.  
This guarantees that the RWZ functions will not be unit-normalized at infinity 
or on the horizon, but instead will have some $A^{\pm}_{lm\o}$ such that 
$X^{\pm}_{lm\o} \sim A^{\pm}_{lm\o} \, e^{\pm i \o r_*}$.  We find 
it advantageous though to construct unit-normalized modes 
$\hat{X}^{\pm}_{lm\o} \sim \exp(\pm i \o r_*)$ \cite{HoppEvan10}.  
To do so we first determine the initial amplitudes $A^{\pm}_{lm\o}$ by 
passing the MST expansions in 
Eqns.~\eqref{eqn:Down1}, \eqref{eqn:DownSeries}, and \eqref{eqn:RMinus} 
through the transformation in Eqns.~\eqref{eqn:RWtrans1} (and additionally 
\eqref{eqn:RWtrans2} as required) to find
\begin{widetext}
\begin{align}
\begin{split}
\label{eqn:RWasymp1}
A_{lm\omega}^{\text{RW},+}
&=
-2^{-1+4 i M \omega} i\omega (M \omega)^{2 i M \omega} 
e^{-\pi  M \omega-\frac{1}{2} i \pi  \nu} \\
&\times\sum_{n=-\infty}^\infty (-1)^n \bigg\{(\nu-1) \nu (\nu+1) (\nu+2)
 + 4 i M \left[2 \nu (\nu+1)-7\right] \omega  +32 i M^3 \omega^3 
+ 400 M^4 \omega^4\\
& \hspace{10ex} + 20 M^2 \left[2 \nu (\nu+1)-1\right] \omega^2 
+ 2 (2 \nu+1) \Big[4 M \omega (5 M \omega+i) + \nu(\nu+1)-1\Big]n \\
&\hspace{10ex} +\Big[8 M \omega (5 M \omega+i)
+ 6 \nu (\nu+1)-1\Big]n^2 
 + (4 \nu+2)n^3 + n^4\bigg\} 
\frac{(\nu-2 i M \omega-1)_n}{(\nu+2 i M \omega+3)_n}a_n, 
\end{split}
\end{align}
\begin{align}
 \begin{split}
\label{eqn:RWasymp2}
A_{lm\omega}^{Z,+}
&= 
-\frac{ 2^{-1+4 i M \omega}i\omega (M\omega)^{2 i M \omega} 
\left[(l-1) l (l+1) (l+2)+12 i M \omega\right]}{(l-1) l (l+1) (l+2)-12 i M 
\omega} \\
&\times\sum_{n=-\infty}^\infty 
e^{\frac{1}{2} i \pi  (2 i M \omega+2 n-\nu)} \Big[2 M \omega (7 i-6 M \omega)
+n(n+2\nu+1)+\nu(\nu+1)\Big] \\
&
\hspace{10ex}
\times \Big\{-2 \left[1+3 M \omega (2 M \omega+i)\right]
 +n(n+2\nu+1)+\nu(\nu+1)\Big\}
\frac{(\nu-2 i M \omega-1)_n}{(\nu+2 i M \omega+3)_n}a_n.
\end{split}
\end{align}	
\begin{align}
\begin{split}
\label{eqn:RWasymp3}
& A_{lm\omega}^{\text{RW},-} 
= A_{lm\omega}^{Z,-} 
= -\frac{1}{M}e^{2 i M \omega} (2 M \omega+i) (4 M \omega+i) 
\sum _{n=-\infty}^{\infty} a_n,
\end{split}
\end{align}
\end{widetext}
These amplitudes are then used to renormalize the initial $a_n$.

\section{Solution to the perturbation equations using MST and SSI}
\label{sec:InhomogSol}

We briefly review here the procedure for solving the perturbation equations 
for eccentric orbits on a Schwarzschild background using MST and a recently 
developed spectral source integration (SSI) \cite{HoppETC15} scheme, both of 
which are needed for high accuracy calculations.  

\subsection{Bound orbits on a Schwarzschild background}
\label{sec:orbits}

We consider generic bound motion between a small mass $\mu$, treated as a 
point particle, and a Schwarzschild black hole of mass $M$, with 
$\mu/M \ll 1$.  Schwarzschild coordinates $x^{\mu} = (t,r,\theta, \varphi )$ 
are used.  The trajectory of the particle is given by 
$x_p^{\a}(\tau) =\left[t_p(\tau),r_p(\tau), \pi/2, \varphi_p(\tau)\right]$ in
terms of proper time $\tau$ (or some other suitable curve parameter) and 
the motion is assumed, without loss of generality, to be confined to the 
equatorial plane.  Throughout this paper, a subscript $p$ denotes evaluation 
at the particle location.  The four-velocity is 
$u^{\alpha} = dx_p^{\alpha}/d\tau$.  

At zeroth order the motion is geodesic in the static background and the 
equations of motion have as constants the specific energy 
$\mathcal{E} = -u_t$ and specific angular momentum $\mathcal{L} = u_\vp$.  
The four-velocity becomes
\be
\label{eqn:four_velocity}
u^\a = \l \frac{{\mathcal{E}}}{f_{p}}, u^r, 0, \frac{{\mathcal{L}}}{r_p^2} \r .
\ee
The constraint on the four-velocity leads to
\be
\label{eqn:rpDots}
\dot r_p^2(t) = f_{p}^2 \left[ 1 - \frac{f_p}{{\mathcal{E}}^2} 
\l 1 + \frac{{\mathcal{L}}^2}{r^2} \r \right] ,
\ee
where dot is the derivative with respect to $t$.  Bound orbits 
have ${\mathcal{E}} < 1$ and, to have two turning points, must at least have 
${\mathcal{L}} > 2 \sqrt{3} M$.  In this case, the pericentric radius, 
$r_{\rm min}$, and apocentric radius, $r_{\rm max}$, serve as alternative 
parameters to ${\mathcal E}$ and ${\mathcal L}$, and also give rise to 
definitions of the (dimensionless) semi-latus rectum $p$ and the 
eccentricity $e$ (see \cite{CutlKennPois94,BaraSago10}).  These various 
parameters are related by 
\be
\label{eqn:defeandp}
{\mathcal{E}}^2 = \frac{(p-2)^2-4e^2}{p(p-3-e^2)},
\q
{\mathcal{L}}^2 = \frac{p^2 M^2}{p-3-e^2} ,
\ee
and $r_{\rm max} = pM/(1-e)$ and $r_{\rm min} = pM/(1+e)$.  The requirement 
of two turning points also sets another inequality, $p > 6 + 2 e$, with the 
boundary $p = 6 + 2 e$ of these innermost stable orbits being the 
separatrix \cite{CutlKennPois94}.

Integration of the orbit is described in terms of an alternate curve 
parameter, the relativistic anomaly $\chi$, that gives the radial position 
a Keplerian-appearing form \cite{Darw59}
\be
r_p \l \chi \r = \frac{pM}{1+ e \cos \chi} .
\ee
One radial libration makes a change $\Delta\chi = 2\pi$.  The orbital 
equations then have the form
\begin{align}
\label{eqn:darwinEqns}
\frac{dt_p}{d \chi} &= \frac{r_p \l \chi \r^2}{M (p - 2 - 2 e \cos \chi)}
 \left[\frac{(p-2)^2 -4 e^2}{p -6 -2 e \cos \chi} \right]^{1/2} ,
\nonumber
\\
\frac{d \varphi_p}{d\chi} 
&= \left[\frac{p}{p - 6 - 2 e \cos \chi}\right]^{1/2} ,
\\
\frac{d\tau_p}{d \chi} &= \frac{M p^{3/2}}{(1 + e \cos \chi)^2} 
\left[ \frac{p - 3 - e^2}{p - 6 - 2 e \cos \chi} \right]^{1/2} ,
\nonumber
\end{align}
and $\chi$ serves to remove singularities in the differential equations
at the radial turning points \cite{CutlKennPois94}.  Integrating the first of 
these equations provides the fundamental frequency and period of radial motion
\be
\label{eqn:O_r}
\O_r \equiv   \frac{2 \pi}{T_r},
\q \q
T_r \equiv \int_{0}^{2 \pi} \l \frac{dt_p}{d\chi} \r d \chi.
\ee
There is an analytic solution to the second equation for the azimuthal 
advance, which is especially useful in our present application,
\be
\vp_p(\chi) = \left(\frac{4 p}{p - 6 - 2 e}\right)^{1/2} \, 
F\left(\frac{\chi}{2} \, \middle| \, -\frac{4 e}{p - 6 - 2 e}  \right) .
\ee
Here $F(x|m)$ is the incomplete elliptic integral of the first kind 
\cite{GradETC07}.  The average of the angular frequency $d \varphi_p / dt$ 
is found by integrating over a complete radial oscillation
\be
\label{eqn:O_phi}
\O_\varphi = \frac{4}{T_r} \left(\frac{p}{p - 6 - 2 e}\right)^{1/2} \, 
K\left(-\frac{4 e}{p - 6 - 2 e}  \right) ,
\ee
where $K(m)$ is the complete elliptic integral of the first kind
\cite{GradETC07}.  Relativistic orbits will have $\Omega_r \ne \Omega_{\vp}$, 
but with the two approaching each other in the Newtonian limit.

\subsection{Solutions to the TD master equation}
\label{sec:TDmasterEq}

This paper draws upon previous work \cite{HoppEvan10} in solving the RWZ 
equations, though here we solve the homogeneous equations using the MST 
analytic function expansions discussed in Sec.~\ref{sec:homog}.  A goal is to 
find solutions to the inhomogeneous time domain (TD) master equations
\be
\label{eqn:masterEqTD}
\left[-\frac{\pa^2}{\pa t^2} + \frac{\pa^2}{\pa r_*^2} - V_l (r) \right]
\Psi_{lm}(t,r) = S_{lm}(t,r) .
\ee
The parity-dependent source terms $S_{lm}$ arise from decomposing the 
stress-energy tensor of a point particle in spherical harmonics.  They are
found to take the form
\begin{align}
\label{eqn:sourceTD}
S_{lm} = G_{lm}(t) \, \delta[r - r_p(t)] + F_{lm}(t) \,
\delta'[r - r_p(t)],
\end{align}
where $G_{lm}(t)$ are $F_{lm}(t)$ are smooth functions.  Because of the 
periodic radial motion, both $\Psi_{lm}$ and $S_{lm}$ can be written as 
Fourier series
\begin{align}
\label{eqn:psiSeries}
\Psi_{lm}(t,r) &= \sum_{n=-\infty}^\infty X_{lmn}(r) \, e^{-i \o t} , \\
S_{lm}(t,r) &= \sum_{n=-\infty}^\infty Z_{lmn}(r) \, e^{-i \o t},
\label{eqn:Slm}
\end{align}
where the $\o \equiv \omega_{mn} = m\Omega_\vp + n\Omega_r$ reflects the 
bi-periodicity of the source motion.  The inverses are
\begin{align}
X_{l mn}(r) &= \frac{1}{T_r} \int_0^{T_r} dt \ \Psi_{l m}(t,r) 
\, e^{i \o t},
\\
Z_{l mn}(r) &= \frac{1}{T_r} \int_0^{T_r} dt \ S_{l m}(t,r) 
\, e^{i \o t} .
\label{eqn:Zlmn}
\end{align}
Inserting these series in Eqn.~\eqref{eqn:masterEqTD} reduces the TD master 
equation to a set of inhomogeneous ordinary differential equations (ODEs) 
tagged additionally by harmonic $n$,
\be
\label{eqn:masterInhomogFD}
\left[\frac{d^2}{dr_*^2} +\omega^2 -V_l (r) \right]
X_{lmn}(r) = Z_{lmn} (r) .
\ee
The homogeneous version of this equation is solved by MST expansions.  The 
unit normalized solutions at infinity (up) are $\hat{X}^+_{lmn}$ while the 
horizon-side (in) solutions are $\hat{X}^-_{lmn}$.  These independent 
solutions provide a Green function, from which the particular solution to 
Eqn.~\eqref{eqn:masterInhomogFD} is derived
\be
\label{eqn:FDInhomog}
X_{lmn} (r) = c^+_{lmn}(r) \, \hat{X}^+_{lmn}(r)
+ c^-_{lmn}(r) \, \hat{X}^-_{lmn}(r) .
\ee
See Ref.~\cite{HoppEvan10} for further details.  However, Gibbs behavior in 
the Fourier series makes reconstruction of $\Psi_{lm}$ in this fashion 
problematic.  Instead, the now standard approach is to derive the TD solution 
using the method of extended homogeneous solutions (EHS) \cite{BaraOriSago08}.

We form first the frequency domain (FD) EHS
\be
\label{eqn:FD_EHS}
X^\pm_{lmn} (r) \equiv C^{\pm}_{lmn} \hat X_{lmn}^\pm (r), \q \q r > 2M ,
\ee
where the normalization coefficients, $C^+_{lmn} = c^+_{lmn}(r_{\rm max})$ and 
$C^-_{lmn} = c^-_{lmn}(r_{\rm min})$, are discussed in the next subsection.
From these solutions we define the TD EHS,
\be
\label{eqn:TD_EHS}
\Psi^\pm_{lm} (t,r) 
\equiv \sum_n X^\pm_{lmn} (r) \, e^{-i \o t}, \q \q r > 2M .
\ee
Then the particular solution to Eqn.~\eqref{eqn:masterEqTD} is formed by 
abutting the two TD EHS at the particle's location,
\begin{align}
\begin{split}
\Psi_{lm} (t,r) &= \Psi^{+}_{lm}(t,r) \theta \left[ r - r_p(t) \right] \\
& \hspace{10ex}
+ 
\Psi^{-}_{lm}(t,r) \theta \left[ r_p(t) - r \right] .
\end{split}
\end{align}

\subsection{Normalization coefficients}
\label{sec:NormCoeff}

The following integral must be evaluated to obtain the normalization 
coefficients $C^\pm_{lmn}$ \cite{HoppEvan10}
\begin{align}
\label{eqn:EHSC}	
C_{lmn}^\pm  
&=   \frac{1}{W_{lmn} T_r} \int_0^{T_r}
\Bigg[ 
 \frac{1}{f_{p}} \hat X^\mp_{lmn}
 G_{lm} \hspace{5ex} \\
&\hspace{5ex} 
+ \l \frac{2M}{r_{p}^2 f_{p}^{2}} \hat X^\mp_{lmn}
 - \frac{1}{f_{p}} 
 \frac{d \hat X^\mp_{lmn}}{dr} \r F_{lm}
 \Bigg]  e^{i \o t}  \, dt, \notag
\end{align}
where $W_{lmn}$ is the Wronskian
\be
W_{lmn} = f \hat{X}^-_{lmn} \frac{d \hat{X}^+_{lmn}}{dr}
- f \hat{X}^+_{lmn} \frac{d \hat{X}^-_{lmn}}{dr} .
\ee
The integral in \eqref{eqn:EHSC} is often computed using Runge-Kutta (or 
similar) numerical integration, which is algebraically convergent.  As shown 
in \cite{HoppETC15} when MST expansions are used with arbitrary-precision 
algorithms to obtain high numerical accuracy (i.e., much higher than double 
precision), algebraically-convergent integration becomes prohibitively 
expensive.  We recently developed the SSI scheme, which provides exponentially 
convergent source integrations, in order to make possible MST calculations of 
eccentric-orbit EMRIs with arbitrary precision.  In the present paper our 
calculations of energy fluxes have up to 200 decimal places of accuracy.

The central idea is that, since the source terms $G_{lm}(t)$ and $F_{lm}(t)$ 
and the modes $X^{\pm}_{lmn}(r)$ are smooth functions, the integrand in 
\eqref{eqn:EHSC} can be replaced by a sum over equally-spaced samples
\be
\label{eqn:CSum}
C^{\pm}_{lmn} = \frac{1}{N W_{lmn}} \sum_{k=0}^{N-1} \bar{E}^{\pm}_{lmn}(t_k) 
\, e^{i n\Omega_r t_k} .
\ee
In this expression $\bar{E}_{lmn}$ is the following $T_r$-periodic
smooth function of time
\begin{align}
&\bar{E}^{\pm}_{lmn}(t) =
\frac{\bar{G}_{lm}(t)}{f_p} \, \hat{X}^{\mp}_{lmn}(r_p(t)) \\
& 
\hspace{5ex}+ 
\frac{2 M}{r_p^2} \frac{\bar{F}_{lm}(t)}{f_p^2} \,\hat{X}^{\mp}_{lmn}(r_p(t)) 
- 
\frac{\bar{F}_{lm}(t)}{f_p} \, \partial_r\hat{X}^{\mp}_{lmn}(r_p(t)) . \notag
\end{align}
It is evaluated at $N$ times that are evenly spaced between $0$ and $T_r$, i.e.,
$t_k \equiv k T_r / N$.  In this expression $\bar{G}_{lm}$ is related to the 
term in Eqn.~\eqref{eqn:sourceTD} by 
$\bar{G}_{lm}=G_{lm} e^{im\Omega_{\vp} t}$ (likewise for $\bar{F}_{lm}$).
It is then found that the sum in \eqref{eqn:CSum} exponentially converges to 
the integral in \eqref{eqn:EHSC} as the sample size $N$ increases.  

One further improvement was found.  The curve parameter in \eqref{eqn:EHSC} 
can be arbitrarily changed and the sum \eqref{eqn:CSum} is thus replaced by
one with even sampling in the new parameter.  Switching from $t$ to $\chi$ 
has the effect of smoothing out the source motion, and as a result the sum
\begin{align}
\label{eqn:CfromEbar}
C^{\pm}_{lmn} &= \frac{\O_r}{N W_{lmn}} \sum_k 
\frac{dt_p}{d\chi}
\bar{E}^{\pm}_{lmn}(t_k)
\, e^{i n\Omega_r t_k} ,
\end{align}
evenly sampled in $\chi$ ($\chi_k = 2 \pi k /N$ with $t_k = t_p(\chi_k)$) 
converges at a substantially faster rate.  This is particularly advantageous 
for computing normalizations for high eccentricity orbits.

Once the $C^{\pm}_{lmn}$ are determined, the energy fluxes at infinity can 
be calculated using
\begin{align}
\label{eqn:fluxNumeric}
\left\langle \frac{dE}{dt} \right\rangle = 
\sum_{lmn}\frac{\o^2}{64\pi}\frac{(l+2)!}{(l-2)!}
|C^{+}_{lmn}|^2 ,
\end{align}
given our initial unit normalization of the modes $\hat{X}_{lmn}^{\pm}$. 
We return to this subject and specific algorithmic details in 
Sec.~\ref{sec:CodeDetails}.

\medskip

\begin{widetext}

\section{Preparing the PN expansion for comparison with perturbation 
theory}
\label{sec:preparePN}

The formalism we briefly discussed in the preceding sections, along with 
the technique in \cite{HoppETC15}, was used to build a code for computing 
energy fluxes at infinity from eccentric orbits to accuracies as high as 200 
decimal places, and to then confirm previous work in PN theory and to discover 
new high PN order terms.  In this section we make further preparation for that 
comparison with PN theory.  The average energy and angular momentum fluxes from 
an eccentric binary are known to 3PN relative order 
\cite{ArunETC08a,ArunETC08b,ArunETC09a} (see also the review by Blanchet 
\cite{Blan14}).  The expressions are given in terms of three parameters; 
e.g., the gauge-invariant post-Newtonian compactness parameter 
$x\equiv\left[(m_1 + m_2) \Omega_\vp\right]^{2/3}$, the eccentricity, and the 
symmetric mass ratio $\nu=m_1 m_2/(m_1+m_2)^2 \simeq\mu/M$ (not to be confused 
with our earlier use of $\nu$ for renormalized angular momentum parameter).  
In this paper we ignore contributions to the flux that are higher order in the 
mass ratio than $\mathcal{O}(\nu^2)$, as these would require a second-order 
GSF calculation to reach.  The more appropriate compactness parameter in the 
extreme mass ratio limit is $y\equiv\left(M \Omega_\vp\right)^{2/3}$, with 
$y = x (1+ m_1/m_2)^{-2/3}$ \cite{Blan14}.  Composed of a set of 
eccentricity-dependent coefficients, the energy flux through 3PN order has 
the form
\begin{align}
\label{eqn:energyflux}
\mathcal{F}_{\rm 3PN} =
\left\langle \frac{dE}{dt} \right\rangle_{\rm 3PN} = 
\frac{32}{5} \left(\frac{\mu}{M}\right)^2 y^5 \, 
\Bigl(\mathcal{I}_0 + y\,\mathcal{I}_1 + y^{3/2}\,\mathcal{K}_{3/2} 
+ y^2\,\mathcal{I}_2 
+ y^{5/2}\,\mathcal{K}_{5/2} 
+ y^3\,\mathcal{I}_3 + y^3\,\mathcal{K}_{3} \Bigr) .
\end{align}
The $\mathcal{I}_n$ are instantaneous flux functions [of eccentricity and 
(potentially) $\log(y)$] that have known closed-form expressions (summarized 
below).  The $\mathcal{K}_n$ coefficients are hereditary, or tail, 
contributions (without apparently closed forms).  The purpose of this 
section is to derive new expansions for these hereditary terms and to 
understand more generally the structure of all of the eccentricity dependent 
coefficients, up to 3PN order and beyond.  

\subsection{Known instantaneous energy flux terms}

For later reference and use, we list here the instantaneous energy flux 
functions, expressed in modified harmonic (MH) gauge 
\cite{ArunETC08a,ArunETC09a,Blan14} and in terms of $e_t$, a particular 
definition of eccentricity (\emph{time eccentricity}) used in the 
quasi-Keplerian (QK) representation \cite{DamoDeru85} of the orbit (see also 
\cite{KoniGopa05,KoniGopa06,ArunETC08a,ArunETC08b,ArunETC09a,GopaScha11,Blan14})
\begin{align}
\label{eqn:edot0} 
\mathcal{I}_0 &= \frac{1}{(1-e_t^2)^{7/2}}
{\left(1+\frac{73}{24}~e_t^2 + \frac{37}{96}~e_t^4\right)} ,
\\
\label{eqn:edot1} 
\mathcal{I}_1 &=
\frac{1}{(1-e_t^2)^{9/2}}
{\left( -\frac{1247}{336} + \frac{10475}{672} e_t^2 + \frac{10043}{384} e_t^4
  + \frac{2179}{1792} e_t^6 \right)} ,
\\
\begin{split}
\label{eqn:edot2} 
\mathcal{I}_2 &= 
\frac{1}{(1-e_t^2)^{11/2}}
{\left(-\frac{203471}{9072} - \frac{3807197}{18144} e_t^2
  - \frac{268447}{24192} e_t^4 + \frac{1307105}{16128} e_t^6
  + \frac{86567}{64512} e_t^8 \right)}
\\
&\hspace{50ex} + \frac{1}{(1-e_t^2)^{5}} \left(\frac{35}{2} 
+ \frac{6425}{48} e_t^2 
  + \frac{5065}{64} e_t^4 + \frac{185}{96} e_t^6 \right) ,
\end{split}
\end{align}
\begin{align}
\begin{split}
\label{eqn:edot3}
\mathcal{I}_3 &= \frac{1}{(1-e_t^2)^{13/2}}
\left(\frac{2193295679}{9979200} + \frac{20506331429}{19958400} e_t^2 
  -\frac{3611354071}{13305600} e_t^4 
\right.
\\
\biggl.
&\hspace{45ex}+ \frac{4786812253}{26611200} e_t^6 
  + \frac{21505140101}{141926400} e_t^8 - \frac{8977637}{11354112} e_t^{10} 
\biggr)
\\&
+ \frac{1}{(1-e_t^2)^{6}} \left(-\frac{14047483}{151200} 
  + \frac{36863231}{100800} e_t^2 + \frac{759524951}{403200} e_t^4 
  + \frac{1399661203}{2419200} e_t^6 + \frac{185}{48} e_t^8 \right) 
\\&
+
\frac{1712}{105}
 \log\left[\frac{y}{y_0}
  \frac{1+\sqrt{1-e_t^2}}{2(1-e_t^2)}\right] F(e_t),
\end{split}
\end{align}
where the function $F(e_t)$ in Eqn.~\eqref{eqn:edot3}
has the following closed-form \cite{ArunETC08a}
\be
F(e_t) = \frac{1}{(1-e_t^2)^{13/2}} 
\bigg( 1+ \frac{85}{6} e_t^2 + \frac{5171}{192} e_t^4 + 
\frac{1751}{192} e_t^6 + \frac{297}{1024} e_t^8\bigg) .
\label{eqn:capFe}
\ee
The first flux function, $\mathcal{I}_0(e_t)$, is the \emph{enhancement 
function} of Peters and Mathews \cite{PeteMath63} that arises from quadrupole 
radiation and is computed using only the Keplerian approximation of the 
orbital motion.  The term ``enhancement function'' is used for functions like 
$\mathcal{I}_0(e_t)$ that are defined to limit on unity as the orbit becomes
circular (with one exception discussed below).  Except for $\mathcal{I}_0$, the 
flux coefficients generally depend upon choice of gauge, compactness 
parameter, and PN definition of eccentricity.  [Note that the extra parameter 
$y_0$ in the $\mathcal{I}_3$ log term cancels a corresponding log term in the 
3PN hereditary flux. See Eqn.~\eqref{eqn:hered3PN} below.]  We also point out 
here the appearance of factors of $1 - e_t^2$ with negative, odd-half-integer 
powers, which make the PN fluxes diverge as $e_t \rightarrow 1$.  We will have 
more to say in what follows about these \emph{eccentricity singular factors.}

\subsection{Making heads or tails of the hereditary terms}

The hereditary contributions to the energy flux can be defined 
\cite{ArunETC08a} in terms of an alternative set of functions
\begin{align}
\label{eqn:edot32}
&\mathcal{K}_{3/2}  = 4\pi\,\varphi(e_t) ,  \\
&\mathcal{K}_{5/2} = -\frac{8191}{672}\,\pi\,\psi(e_t) 
 \label{eqn:hered52} , \\
&\mathcal{K}_{3} = -\frac{1712}{105}\,\chi(e_t) +
\left[
-\frac{116761}{3675} + \frac{16}{3} \,\pi^2 -\frac{1712}{105}\,\gamma_\text{E} -
  \frac{1712}{105}\log\left(\frac{4y^{3/2}}{y_0}\right)\right]\, F(e_t) ,
\label{eqn:hered3PN}
\end{align}
where $\gamma_\text{E}$ is the Euler constant and $F$, $\vp$, $\psi$, and 
$\chi$ are enhancement functions (though $\chi$ is the aforementioned special
case, which instead of limiting on unity vanishes as $e_t \rightarrow 0$).  
(Note also that the enhancement function $\chi(e_t)$ should not to be 
confused with the orbital motion parameter $\chi$.)  Given the limiting 
behavior of these new functions, the circular orbit limit 
becomes obvious.  The 1.5PN enhancement function $\vp$ was first calculated 
by Blanchet and Sch\"{a}fer \cite{BlanScha93} following discovery of the 
circular orbit limit ($4 \pi$) of the tail by Wiseman \cite{Wise93} 
(analytically) and Poisson \cite{Pois93} (numerically, in an early BHP 
calculation).  The function $F(e_t)$, given above in Eqn.~\eqref{eqn:capFe},
is closed form,
while $\vp$, $\psi$, and $\chi$ (apparently) are not.  Indeed, the lack of 
closed-form expressions for $\vp$, $\psi$, and $\chi$ presented a problem 
for us.  Arun et al.~\cite{ArunETC08a,ArunETC08b,ArunETC09a} computed these 
functions numerically and plotted them, but gave only low-order expansions in 
eccentricity.  For example Ref.~\cite{ArunETC09a} gives for the 1.5PN tail 
function
\be
\label{eqn:arunphi}
\vp(e_t)=1+\frac{2335}{192} e_t^2 + \frac{42955}{768} e_t^4 + \cdots .  
\ee
One of the goals of this paper became finding means of calculating these 
functions with (near) arbitrary accuracy.

The expressions above are written as functions of the eccentricity $e_t$.  
However, the 1.5PN tail $\vp$ and the functions $F$ and $\chi$ only depend 
upon the binary motion, and moments, computed to Newtonian order.  Hence, for 
these functions (as well as $\mathcal{I}_0$) there is no distinction between 
$e_t$ and the usual Keplerian eccentricity.  Nevertheless, since we will 
reserve $e$ to denote the relativistic (Darwin) eccentricity, we express 
everything here in terms of $e_t$.

Blanchet and Sch\"{a}fer \cite{BlanScha93} showed that $\vp(e_t)$, like the 
Peters-Mathews enhancement function $\mathcal{I}_0$, is determined by the 
quadrupole moment as computed at Newtonian order from the Keplerian elliptical 
motion.  Using the Fourier series expansion of the time dependence of a Kepler
ellipse \cite{PeteMath63,Magg07}, $\mathcal{I}_0$ can be written in 
terms of Fourier amplitudes of the quadrupole moment by 
\be
\label{eqn:pmsum}
\mathcal{I}_0(e_t) = \frac{1}{16} \sum_{n=1}^\infty n^6 
\vert \! \mathop{\hat{I}}_{(n)}{}_{\!\!ij}^{\!\!(\mathrm{N})} \vert^2 = 
\sum_{n=1}^\infty g(n,e_t) = f(e_t) =
\frac{1}{(1-e_t^2)^{7/2}}
{\left(1+\frac{73}{24}~e_t^2 + \frac{37}{96}~e_t^4\right)} ,
\ee
which is the previously mentioned closed form expression.  Here, $f(e)$ is 
the traditional Peters-Mathews function name, which is not to be confused 
with the metric function $f(r)$.  In the expression, 
${}_{(n)} \hat{I}_{ij}^{(\mathrm{N})}$ is the $n$th Fourier harmonic of the 
dimensionless quadrupole moment (see sections III through V of 
\cite{ArunETC08a}).  The function $g(n,e_t)$ that represents the square of 
the quadrupole moment amplitudes is given by
\begin{align}
\label{eqn:gfunc}
g(n,e_t) &\equiv \frac{1}{2} n^2 \bigg\{ \left[-\frac{4}{e_t^3}-3 e_t+
\frac{7}{e_t}\right] n J_n(n e_t) J_n'(n e_t) +
\left[\left(e_t^2+\frac{1}{e_t^2}-2\right) n^2+\frac{1}{e_t^2}-1\right] 
J_n'(n e_t)^2 \notag\\ 
& \hspace{40ex} +\left[\frac{1}{e_t^4}-\frac{1}{e_t^2}+
\left(\frac{1}{e_t^4}-e_t^2-\frac{3}{e_t^2}+3\right) n^2+\frac{1}{3}\right] 
 J_n(ne_t)^2\bigg\} ,
\end{align}
and was derived by Peters and Mathews \cite{PeteMath63} (though the corrected 
expression can be found in \cite{BlanScha93} or \cite{Magg07}).  

These quadrupole moment amplitudes also determine $F(e_t)$,
\be
\label{eqn:capFeSum}
F(e_t) = \frac{1}{4} \sum_{n=1}^\infty n^2 \, g(n,e_t) ,
\ee
whose closed form expression is found in \eqref{eqn:capFe}, and the 1.5PN 
tail function \cite{BlanScha93}, which emerges from a very similar sum
\be
\label{eqn:phi2}
\vp(e_t) = \sum_{n=1}^\infty \frac{n}{2} \, g(n,e_t) .
\ee
Unfortunately, the odd factor of $n$ in this latter sum (and more generally 
any other odd power of $n$) makes it impossible to translate the sum into an 
integral in the time domain and blocks the usual route to finding a closed-form 
expression like $f(e_t)$ and $F(e_t)$.

The sum \eqref{eqn:phi2} might be computed numerically but it is more 
convenient to have an expression that can be understood at a glance and be 
rapidly evaluated.  The route we found to such an expression leads to 
several others.  We begin with \eqref{eqn:gfunc} and expand $g(n,e_t)$, 
pulling forward the leading factor and writing the remainder as
a Maclaurin series in $e_t$
\be
\label{eqn:gexp}
g(n,e_t) = \left(\frac{n}{2}\right)^{2n} e_t^{2n - 4} \left(
\frac{1}{\Gamma(n-1)^2} - 
\frac{(n-1)(n^2 + 4n -2)}{2 \, \Gamma(n)^2} e_t^2 + 
\frac{6 n^4 + 45 n^3 + 18 n^2 - 48 n + 8}{48 \, \Gamma(n)^2} e_t^4 +
\cdots \right) .
\ee
In a sum over $n$, successive harmonics each contribute a series that starts 
at a progressively higher power of $e_t^2$.  Inspection further shows that for 
$n = 1$ the $e_t^{-2}$ and $e_t^0$ terms vanish, the former because 
$\Gamma(0)^{-1} \rightarrow 0$.  The $n = 2$ harmonic is the only one that 
contributes at $e_t^0$ [in fact giving $g(2,e_t) = 1$, the circular orbit 
limit].  The successively higher-order power series in $e_t^2$ imply that the 
individual sums that result from expanding \eqref{eqn:pmsum}, 
\eqref{eqn:capFeSum}, and \eqref{eqn:phi2} each truncate, with only a finite 
number of harmonics contributing to the coefficient of any given power of 
$e_t^2$.

If we use \eqref{eqn:gexp} in \eqref{eqn:pmsum} and sum, we find 
$\mathcal{I}_0 = 1 + (157/24) e_t^2 + (605/32) e_t^4 + (3815/96) e_t^6 + 
\cdots$, an infinite series.  If on the other hand we introduce the known 
eccentricity singular factor, take $(1 - e_t^2)^{7/2} \, g(n,e_t)$, re-expand 
and sum, we then find $1 + (73/24) e_t^2 + (37/96) e_t^4$, the well known 
Peters-Mathews polynomial term.  All the sums for higher-order terms vanish 
identically.  The same occurs if we take a different eccentricity singular 
factor, expand $(1/4) (1 - e_t^2)^{13/2} \, n^2 \, g(n,e_t)$ and sum; we 
obtain the polynomial in the expression for $F(e_t)$ found in 
\eqref{eqn:capFe}.  The power series expansion of $g(n,e_t)$ thus provides 
an alternative means of deriving these enhancement functions without 
transforming to the time domain.

\subsubsection{Form of the 1.5PN Tail}
\label{sec:tailvarphi}

Armed with this result, we then use \eqref{eqn:gexp} in \eqref{eqn:phi2} and 
calculate the sums in the expansion, finding 
\be
\vp(e_t)=1+ \frac{2335}{192} e_t^2 + \frac{42955}{768} e_t^4 + 
\frac{6204647}{36864} e_t^6 + 
\frac{352891481}{884736} e_t^8 + \cdots ,
\ee
agreeing with and extending the expansion \eqref{eqn:arunphi} derived by 
Arun et al \cite{ArunETC09a}.  We forgo giving a lengthier expression because 
a better form exists.  Rather, we introduce an assumed singular factor and 
expand $(1-e_t^2)^5 \, g(n,e_t)$.  Upon summing we find  
\begin{align}
\label{eqn:vphiExpand}
\varphi(e_t) = &\frac{1}{(1-e_t^2)^5} \, 
\bigg(1+\frac{1375}{192} e_t^2+\frac{3935}{768} e_t^4 
+\frac{10007}{36864} e_t^6 
+\frac{2321}{884736} e_t^8  
+\frac{237857}{353894400} e_t^{10}
+\frac{182863}{4246732800} e_t^{12} \\ \notag
&+\frac{4987211}{6658877030400} e_t^{14} 
-\frac{47839147}{35514010828800} e_t^{16} 
-\frac{78500751181}{276156948204748800} e_t^{18} 
-\frac{3031329241219}{82847084461424640000} e_t^{20} 
+\cdots \bigg) .
\end{align}
Only the leading terms are shown here; we have calculated over 100 terms with 
\emph{Mathematica} and presented part of this expansion previously (available 
online \cite{Fors14,Fors15,Evan15a}).  The first four terms are also published 
in \cite{SagoFuji15}.  The assumed singular factor turns out to be the correct 
one, allowing the remaining power series to converge to a finite value at 
$e_t = 1$.  As can be seen from the rapidly diminishing size of higher-order 
terms, the series is convergent.  The choice for singular factor is supported 
by asymptotic analysis found in Sec.~\ref{sec:asymptotic}.  The 1.5PN 
singular factor and the high-order expansion of $\vp(e_t)$ are two key 
results of this paper.  

The singular behavior of $\vp(e_t)$ as $e_t\rightarrow 1$ is evident on the 
left in Fig.~\ref{fig:curlyPhiPlot}.  The left side of this figure reproduces 
Figure 1 of Blanchet and Sch\"{a}fer \cite{BlanScha93}, though note their 
older definition of $\vp(e)$ (Figure 1 of Ref.~\cite{ArunETC08a} compares 
directly to our plot).  The right side of Fig.~\ref{fig:curlyPhiPlot} however 
shows the effect of removing the singular dependence and plotting only the 
convergent power series.  We find that the resulting series limits on 
$\simeq 13.5586$ at $e_t = 1$.

\begin{figure}
\includegraphics[scale=.95]{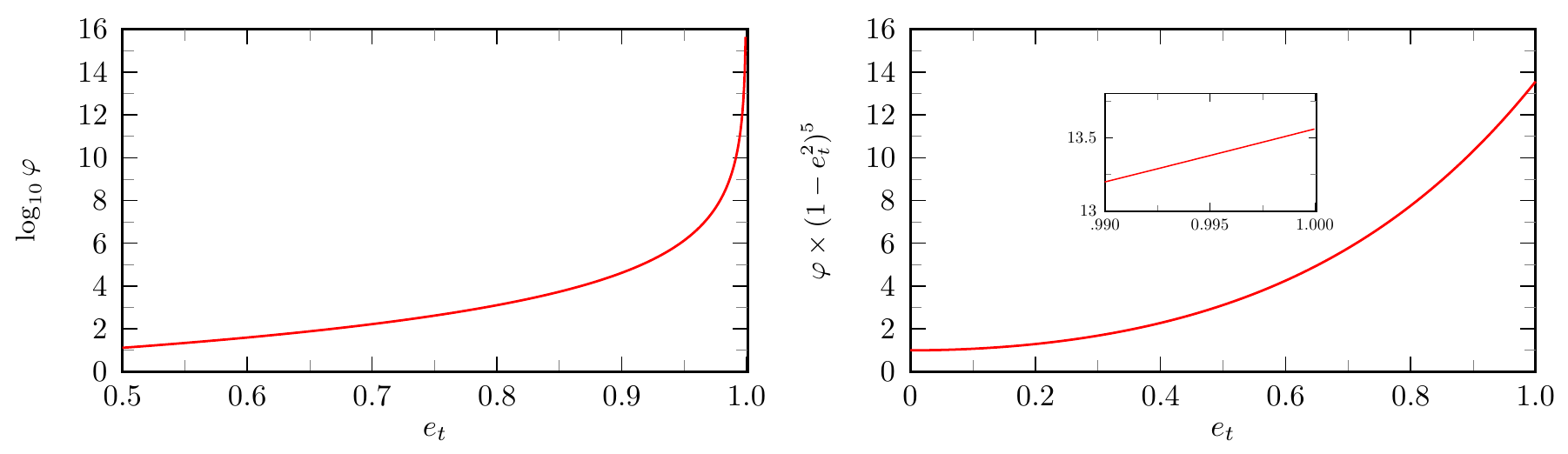}
\caption{
Enhancement function $\varphi(e_t)$ associated with the 1.5PN tail.  On the 
left the enhancement function is directly plotted, demonstrating the singular 
behavior as $e_t \rightarrow 1$.  On the right, the eccentricity singular 
factor $(1-e_t^2)^{-5}$ is removed to reveal convergence in the remaining 
expansion to a finite value of approximately $13.5586$ at $e_t = 1$.  
\label{fig:curlyPhiPlot}} 
\end{figure}

\subsubsection{Form of the 3PN Hereditary Terms}
\label{sec:chiform}

With a useful expansion of $\vp(e_t)$ in hand, we employ the same approach to 
the other hereditary terms.  As a careful reading of Ref.~\cite{ArunETC08a} 
makes clear the most difficult contribution to calculate is 
\eqref{eqn:hered52}, the correction of the 1.5PN tail showing up at 2.5PN 
order.  Accordingly, we first consider the simpler 3PN case 
\eqref{eqn:hered3PN}, which is the sum of the tail-of-the-tail and 
tail-squared terms \cite{ArunETC08a}.  The part in \eqref{eqn:hered3PN} that 
requires further investigation is $\chi(e_t)$.  The infinite series for 
$\chi(e_t)$ is shown in \cite{ArunETC08a} to be
\be
\label{eqn:chiSum}
\chi(e_t) = \frac{1}{4} \, \sum_{n=1}^\infty n^2 \, 
\log\left(\frac{n}{2}\right) \, g(n,e_t) .
\ee
The same technique as before is now applied to $\chi(e_t)$ using the expansion 
\eqref{eqn:gexp} of $g(n,e_t)$.  The series will be singular at $e_t = 1$, so 
factoring out the singular behavior is important.  However, for reasons to 
be explained in Sec.~\ref{sec:asymptotic}, it proves essential in this case 
to remove the two strongest divergences.  We find 
\begin{align}
\label{eqn:chiExp}
\chi (e_t) = - \frac{3}{2} F(e_t) \log(1 - e_t^2) &+
\frac{1}{(1-e_t^2)^{13/2}} \, 
\Bigg\{ 
\left[-\frac{3}{2} - \frac{77}{3} \log(2) + \frac{6561}{256} \log(3) \right] 
e_t^2 \\ \notag
&+
\left[-22 + \frac{34855}{64} \log(2) - \frac{295245}{1024} \log(3) \right] 
e_t^4 
\\ \notag
&+
\left[-\frac{6595}{128} - \frac{1167467}{192} \log(2) + 
\frac{24247269}{16384} \log(3) + \frac{244140625}{147456} \log(5) \right] 
e_t^6 
\\ \notag
&+
\left[-\frac{31747}{768} + \frac{122348557}{3072} \log(2) + 
\frac{486841509}{131072} \log(3) - \frac{23193359375}{1179648} \log(5) \right] 
e_t^8 
+ \cdots 
\Bigg\} . 
\end{align}
Empirically, we found the series for $\chi(e_t)$ diverging like
$\chi(e_t) \sim - C_{\chi} (1 - e_t^2)^{-13/2} \log(1 - e_t^2)$ as 
$e_t \rightarrow 1$, where $C_{\chi}$ is a constant.  The first term in 
\eqref{eqn:chiExp} apparently encapsulates all of the logarithmic divergence 
and implies that $C_{\chi} = -(3/2)(52745/1024) \simeq -77.2632$.  The reason 
for pulling out this particular function is based on a guess suggested by the 
asymptotic analysis in Sec.~\ref{sec:asymptotic} and considerations on how 
logarithmically divergent terms in the combined instantaneous-plus-hereditary 
3PN flux should cancel when a switch is made from orbital parameters 
$e_t$ and $y$ to parameters $e_t$ and $1/p$ (to be further discussed in a 
forthcoming paper).  Having isolated the two divergent terms, the remaining 
series converges rapidly with $n$.  The divergent behavior of the second term
as $e_t \rightarrow 1$ is computed to be approximately 
$\simeq +73.6036 (1-e_t^2)^{-13/2}$.  The appearance of $\chi(e_t)$ is shown 
in Fig.~\ref{fig:chiPlot}, with and without its most singular factor removed. 

\subsubsection{Form of the 2.5PN Hereditary Term}

Armed with this success we went hunting for a comparable result for the 
2.5PN enhancement factor $\psi$.  Calculating $\psi$ is a much more involved 
process, as part of the tail at this order is a 1PN correction to the mass 
quadrupole.  At 1PN order the orbital motion no longer closes and the 
corrections in the mass quadrupole moments require a biperiodic Fourier 
expansion.  Arun et al.~\cite{ArunETC08b} describe a procedure for computing 
$\psi$, which they evaluated numerically.  One of the successes we are able 
to report in this paper is having obtained a high-order power series 
expansion for $\psi$ in $e_t$.  Even with \emph{Mathematica's} help, it is 
a consuming calculation, and we have reached only the 35th order 
($e_t^{70}$).  This achieves some of our purposes in seeking the expansion.
We were also able to predict the comparable singular factor present as 
$e_t \rightarrow 1$ and demonstrate apparent convergence in the remaining 
series to a finite value at $e_t = 1$.  The route we followed in making the 
calculation of the 2.5PN tail is described in App.~\ref{sec:massQuad}.  
Here, we give the first few terms in the $\psi$ expansion
\begin{align}
\label{eqn:psiExpand}
\psi(e_t)  =  \frac{1}{\left(1-e_t^2\right)^6}
\biggl[
1 &-\frac{72134}{8191}e_t^2-\frac{19817891}{524224}e_t^4
-\frac{62900483}{4718016}e_t^6-\frac{184577393}{603906048}e_t^8
+\frac{1052581}{419379200}e_t^{10} \notag \\
&-\frac{686351417}{1159499612160}e_t^{12} 
+\frac{106760742311}{852232214937600}e_t^{14}
+\frac{7574993235161}{436342894048051200}e_t^{16} \notag \\
&-\frac{4345876114169}{2524555315563724800}e_t^{18} 
-\frac{61259745206138959}{56550039068627435520000}e_t^{20}
+\cdots
 \biggr], 
\end{align}
Like in the preceding plots, we show $\log_{10}|\psi|$ graphed on the left in 
Fig.~\ref{fig:psiPlot}.  The singular behavior is evident.  On the right 
side, the 2.5PN singular factor has been removed and the finite limit at 
$e_t = 1$ is clear.

\begin{figure*}
\includegraphics[scale=.95]{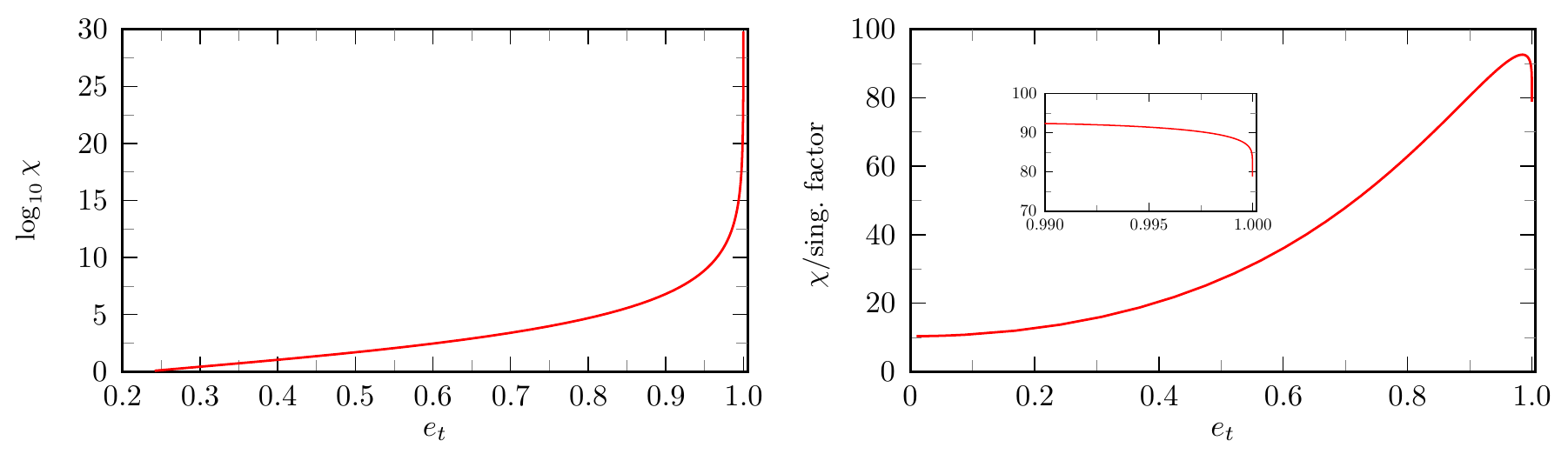}
\caption{
The 3PN enhancement function $\chi(e_t)$.  Its log is plotted on the left.  
On the right we remove the dominant singular factor 
$-(1-e_t^2)^{-13/2} \log(1-e_t^2)$.  The turnover near $e_t = 1$ reflects 
competition with the next-most-singular factor, $(1-e_t^2)^{-13/2}$.
\label{fig:chiPlot}} 
\end{figure*}

\begin{figure*}
\includegraphics[scale=.95]{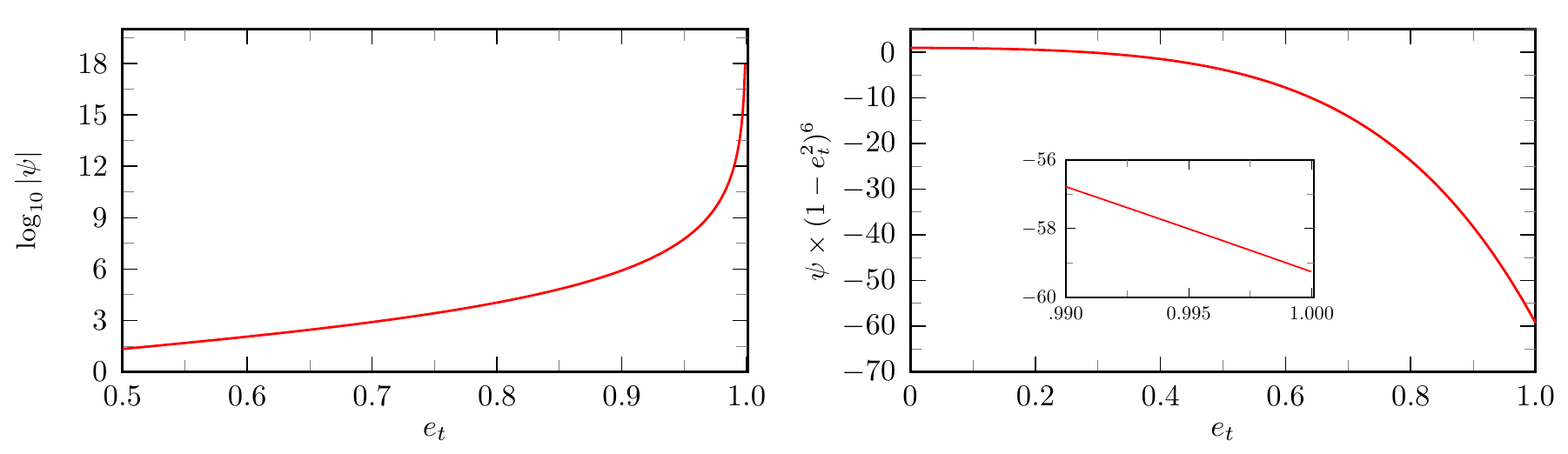}
\caption{
The enhancement factor $\psi(e_t)$.  On the right we remove the singular 
factor $(1-e_t^2)^{-6}$ and see the remaining contribution smoothly approach 
a finite value at $e_t=1$.
\label{fig:psiPlot}} 
\end{figure*}

\subsection{Applying asymptotic analysis to determine eccentricity singular 
factors}
\label{sec:asymptotic}

In the preceding section we assumed the existence of certain ``correct'' 
eccentricity singular factors in the behavior of the known hereditary 
terms, which once factored out allow the remaining power series to converge 
to constants at $e_t = 1$.  We show now that at least some of these 
singular factors, specifically the ones associated with $\varphi(e_t)$ and 
$\chi(e_t)$, can be derived via asymptotic analysis.  In the process the 
same analysis confirms the singular factors in $f(e_t)$ and $F(e_t)$ already 
known from post-Newtonian work.  \emph{As a bonus our asymptotic analysis can 
even be used to make remarkably sharp estimates of the limiting constant 
coefficients that multiply these singular factors.}

What all four of these enhancement functions share is dependence on the 
square of the harmonics of the quadrupole moment given by the function 
$g(n,e_t)$ found in \eqref{eqn:gfunc}.  To aid our analysis near $e_t = 1$, 
we define $x \equiv 1 - e_t^2$ and use $x$ to rewrite \eqref{eqn:gfunc} as
\be
\label{eqn:gxfunc}
g(n,e_t) = 
\frac{1}{6} n^2 
\frac{1+x+x^2 + 3 n^2 x^3}{(1-x)^2} J_n(ne_t)^2 
+\frac{1}{2} n^2 \frac{x (1 + n^2 x)}{1-x} J_n'(n e_t)^2 
-\frac{1}{2} n^3 \frac{x (1 + 3x)}{(1-x)^{3/2}} J_n(n e_t) J_n'(n e_t) .
\ee
An inspection of how \eqref{eqn:gxfunc} folds into \eqref{eqn:pmsum}, 
\eqref{eqn:capFeSum}, \eqref{eqn:phi2}, and \eqref{eqn:chiSum} shows that 
infinite sums of the following forms are required to compute $\varphi(e_t)$, 
$\chi(e_t)$, $f(e_t)$, and $F(e_t)$ 
\be
\label{eqn:Hfunctions}
H_0^{\alpha,\beta} = \sum_{n=1}^{\infty} n^\alpha \log^\beta\left(\frac{n}{2}
\right) 
J_n(n e_t)^2 , \quad
H_1^{\alpha,\beta} = \sum_{n=1}^{\infty} n^\alpha \log^\beta\left(\frac{n}{2}
\right) 
J_n^\prime (n e_t)^2 , \quad
H_2^{\alpha,\beta} = \sum_{n=1}^{\infty} n^\alpha \log^\beta\left(\frac{n}{2}
\right) 
J_n(n e_t) J_n^\prime(n e_t) .
\ee
In this compact shorthand, $\beta=1$ merely indicates sums that contain logs 
needed to calculate $\chi(e_t)$ while $\beta=0$ (absence of a log) covers 
the other cases.  Careful inspection of \eqref{eqn:gxfunc} reveals there are 
18 different sums needed to calculate the four enhancement functions in 
question, and $\alpha$ ranges over (some) values between $2$ and $6$.  

As $x \rightarrow 0$ ($e_t \rightarrow 1$) large $n$ terms have growing 
importance in the sums.  In this limit the Bessel functions have uniform 
asymptotic expansions for large order $n$ of the form 
\cite{AbraSteg72,DLMF,NHMF}
\begin{align}
J_n(n e_t) \sim & \left(\frac{4 \zeta}{x} \right)^{\frac{1}{4}} 
\left[n^{-1/3} {\rm Ai}(n^{2/3} \zeta) 
\sum_{k=0}^{\infty} 
\frac{A_k}{n^{2 k}} + n^{-5/3} {\rm Ai^\prime} (n^{2/3} \zeta) 
\sum_{k=0}^{\infty} 
\frac{B_k}{n^{2 k}} 
\right] , \\
J_n^\prime (n e_t) \sim & -\frac{2}{\sqrt{1-x}} 
\left(\frac{x}{4 \zeta} \right)^{\frac{1}{4}} 
\left[n^{-4/3} {\rm Ai}(n^{2/3} \zeta) 
\sum_{k=0}^{\infty} 
\frac{C_k}{n^{2 k}} + n^{-2/3} {\rm Ai^\prime} (n^{2/3} \zeta) 
\sum_{k=0}^{\infty} 
\frac{D_k}{n^{2 k}} 
\right] , 
\end{align}
where $\zeta$ depends on eccentricity and is found from 
\be
\frac{2}{3} \zeta^{3/2} = \log\left(\frac{1 + \sqrt{x}}{\sqrt{1-x}} \right) 
- \sqrt{x} \equiv \rho(x) \simeq 
\frac{1}{3} x^{3/2} + \frac{1}{5} x^{5/2} + \frac{1}{7} x^{7/2} + \cdots ,
\ee
and where the expansion of $\rho(x)$ is the Puiseux series.  Defining 
$\xi \equiv n \rho(x)$, we need in turn the asymptotic expansions of the 
Airy functions \cite{AbraSteg72,DLMF,NHMF}
\begin{align}
{\rm Ai}(n^{2/3} \zeta)  \sim & \,
\frac{e^{-\xi}}{2^{5/6} 3^{1/6} \sqrt{\pi} \xi^{1/6}} \left(
1 - \frac{5}{72 \xi} + \frac{385}{10368 \xi^2} 
-\frac{85085}{2239488 \xi^3} 
+\frac{37182145}{644972544 \xi^4}
-\frac{5391411025}{46438023168 \xi^5} 
+\cdots  \right) ,
\\
{\rm Ai^\prime} (n^{2/3} \zeta) \sim & \, 
-\frac{3^{1/6} \xi^{1/6} e^{-\xi}}{2^{7/6} \sqrt{\pi}} \left(
1 + \frac{7}{72 \xi} - \frac{455}{10368 \xi^2} 
+\frac{95095}{2239488 \xi^3} 
-\frac{40415375}{644972544 \xi^4}
+\frac{5763232475}{46438023168 \xi^5} 
+\cdots  \right) .
\end{align}
In some of the following estimates all six leading terms in the Airy 
function expansions are important, while a careful analysis reveals that we 
never need to retain any terms in the Bessel function expansions beyond
$A_0 = 1$ and $D_0 = 1$. 

These asymptotic expansions can now be used to analyze the behavior of the 
sums in \eqref{eqn:Hfunctions} (from whence follow the enhancement functions) 
in the limit as $e_t \rightarrow 1$.  Take as an example $H_2^{3,0}$.  We 
replace the Bessel functions with their asymptotic expansions and thus 
obtain an approximation for the sum 
\be
\label{eqn:aeH230}
H_2^{3,0} = \sum_{n=1}^{\infty} n^3 J_n(n e_t) J_n^\prime(n e_t) \simeq
\frac{1}{2 \pi \sqrt{1-x}} \sum_{n=1}^{\infty} n^2 e^{-2 \xi} 
\left(1 + \frac{1}{36 \xi} - \frac{35}{2592 \xi^2} + \cdots \right) ,
\ee
where recall that $\xi$ is the product of $n$ with $\rho(x)$.  The original 
sum has in fact a closed form that can be found in the appendix of 
\cite{PeteMath63}
\be
\sum_{n=1}^{\infty} n^3 J_n(n e_t) J_n^\prime(n e_t) =
\frac{e_t}{4 (1-e_t^2)^{9/2}}
\left(1+3~e_t^2 + \frac{3}{8}~e_t^4\right) \sim
\frac{35}{32} \, \frac{1}{(1-e_t^2)^{9/2}} \simeq 
\frac{1.094}{(1-e_t^2)^{9/2}} ,
\ee
where in the latter part of this line we give the behavior near $e_t =1$.
With this as a target, we take the approximate sum in \eqref{eqn:aeH230} and 
make a further approximation by replacing the sum over $n$ with an integral 
over $\xi$ from $0$ to $\infty$ while letting 
$\Delta n = 1 \rightarrow d\xi / \rho(x)$ and retaining only terms in the 
expansion that yield non-divergent integrals.  We find
\be
\frac{1}{2 \pi \sqrt{1-x}} \frac{1}{\rho(x)^3} 
\int_{0}^{\infty} d\xi \, e^{-2 \xi} \,
\left(\xi^2 + \frac{1}{36} \xi - \frac{35}{2592} \right) = 
\frac{1297}{10368 \, \pi} \frac{1}{\rho(x)^3 \sqrt{1-x}} \sim
\frac{1297}{384 \, \pi} \frac{1}{(1-e_t^2)^{9/2}} \simeq 
\frac{1.0751}{(1-e_t^2)^{9/2}} ,
\ee
with the final result coming from further expanding in powers of $x$.  Our 
asymptotic calculation, and approximate replacement of sum with integral, 
not only provides the known singular dependence but also an estimate of the 
coefficient on the singular term that is better than we perhaps had any 
reason to expect.

All of the remaining 17 sums in \eqref{eqn:Hfunctions} can be approximated 
in the same way.  As an aside it is worth noting that for those sums in 
\eqref{eqn:Hfunctions} without log terms (i.e., $\beta = 0$) the replacement of 
the Bessel functions with their asymptotic expansions leads to infinite sums 
that can be identified as the known polylogarithm functions \cite{DLMF,NHMF}
\be
{\rm Li}_{-k}\left( e^{-2\rho(x)}\right) 
= \sum_{n=1}^{\infty} n^k e^{-2 n \rho(x)} .
\ee
However, expanding the polylogarithms as $x \rightarrow 0$ provides results 
for the leading singular dependence that are no different from those of the 
integral approximation.  Since the $\beta = 1$ cases are not represented by 
polylogarithms, we simply uniformly use the integral approximation.  

We can apply these estimates to the four enhancement functions.  First, the 
Peters-Mathews function $f(e_t)$ in \eqref{eqn:pmsum} has known leading 
singular dependence of
\be
f(e_t) \simeq  \left(1+\frac{73}{24} + \frac{37}{96}\right)
\frac{1}{(1-e_t^2)^{7/2}} = \frac{425}{96} \frac{1}{(1-e_t^2)^{7/2}} \simeq 
\frac{4.4271}{(1-e_t^2)^{7/2}} , \qquad {\rm as} \qquad e_t \rightarrow 1 .
\ee
If we instead make an asymptotic analysis of the sum in \eqref{eqn:pmsum} we 
find
\be
f(e_t) \sim \frac{191755}{13824 \, \pi} \frac{1}{(1-e_t^2)^{7/2}} \simeq 
\frac{4.4153}{(1-e_t^2)^{7/2}} ,
\ee
which extracts the correct eccentricity singular function and yields a 
surprisingly sharp estimate of the coefficient.  We next turn to the function
$F(e_t)$ in \eqref{eqn:capFeSum}.  In this case the function tends to 
$F(e_t) \simeq (52745/1024) (1-e_t^2)^{-13/2} \simeq 51.509 (1-e_t^2)^{-13/2}$
as $e_t \rightarrow 1$.  Using instead the asymptotic technique we get an
estimate 
\be
F(e_t) \sim \frac{5148642773}{31850496 \,\pi} \frac{1}{(1-e_t^2)^{13/2}} \simeq 
\frac{51.455}{(1-e_t^2)^{13/2}} .
\ee
Once again the correct singular function emerges and a surprisingly accurate 
estimate of the coefficient is obtained.  

These two cases are heartening checks on the asymptotic analysis but of 
course both functions already have known closed forms.  What is more 
interesting is to apply the approach to $\varphi(e_t)$ and $\chi(e_t)$, which 
are not known analytically.  For the sum in \eqref{eqn:phi2} for 
$\varphi(e_t)$ we obtain the following asymptotic estimate
\be
\varphi(e_t) \sim 
\frac{56622073}{1327104 \, \pi} \frac{1}{(1-e_t^2)^{5}} -
\frac{371833517}{6635520 \, \pi} \frac{1}{(1-e_t^2)^{4}} + \cdots
\simeq 
\frac{13.581}{(1-e_t^2)^{5}} -
\frac{17.837}{(1-e_t^2)^{4}} + \cdots ,
\ee
where in this case we retained the first two terms in the expansion about 
$e_t = 1$.  The leading singular factor is exactly the one we identified 
in \ref{sec:tailvarphi} and its coefficient is remarkably close to 
the 13.5586 value found by numerically evaluating the high-order expansion in 
\eqref{eqn:vphiExpand}.  The second term was retained merely to illustrate 
that the expansion is a regular power series in $x$ starting with $x^{-5}$ 
(in contrast to the next case).  

We come finally to the enhancement function, $\chi(e_t)$, whose definition 
\eqref{eqn:chiSum} involves logarithms.  Using the same asymptotic 
expansions and integral approximation for the sum, and retaining the 
first two divergent terms, we find
\be
\label{eqn:chiasymp}
\chi(e_t) \sim 
- \frac{5148642773}{21233664 \, \pi} \frac{\log(1-e_t^2)}{(1-e_t^2)^{13/2}} 
- \frac{5148642773}{21233664 \, \pi} \left[-\frac{7882453164}{5148642773} 
+ \frac{2}{3} \gamma_E + \frac{4}{3} \log (2) 
- \frac{2}{3} \log(3) \right] \frac{1}{(1-e_t^2)^{13/2}} .
\ee
The form of \eqref{eqn:chiExp} assumed in Sec.~\ref{sec:chiform}, whose 
usefulness was verified through direct high-order expansion, was suggested 
by the leading singular behavior emerging from this asymptotic analysis. 
We guessed that there would be two terms, one with eccentricity singular 
factor $\log(1-e_t^2) (1-e_t^2)^{-13/2}$ and one with $(1-e_t^2)^{-13/2}$. 
In any calculation made close to $e_t = 1$ these two leading terms compete 
with each other, with the logarithmic term only winning out slowly as 
$e_t \rightarrow 1$.  Prior to identifying the two divergent series we 
initially had difficulty with slow convergence of an expansion for 
$\chi(e_t)$ in which only the divergent term with the logarithm was factored 
out.  To see the issue, it is useful to numerically evaluate our approximation 
\eqref{eqn:chiasymp} 
\be
\label{eqn:chiasympnum}
\chi(e_t) \sim -77.1823 \frac{\log(1-e_t^2)}{(1-e_t^2)^{13/2}} 
\left[1 - \frac{0.954378}{\log(1-e_t^2)} \right] = 
-77.1823 \frac{\log(1-e_t^2)}{(1-e_t^2)^{13/2}} 
+ \frac{73.6612}{(1-e_t^2)^{13/2}} .
\ee
From this it is clear that even at $e_t = 0.99$ the second term makes a 
$+24.3$\% correction to the first term, giving the misleading impression 
that the leading coefficient is near $-96$ not $-77$.  The key additional 
insight was to guess the closed form for the leading singular term in 
\eqref{eqn:chiExp}.  As mentioned, the reason for expecting this exact 
relationship comes from balancing and cancelling logarithmic terms in both 
instantaneous and hereditary 3PN terms when the expansion is converted
from one in $e_t$ and $y$ to one in $e_t$ and $1/p$.  The coefficient on 
the leading (logarithmic) divergent term in $\chi(e_t)$ is exactly 
$-(3/2)(52745/1024) \simeq -77.2632$.  [This number is -3/2 times the limit 
of the polynomial in $F(e_t)$.]  It compares well with the first number in 
\eqref{eqn:chiasympnum}.  Additionally, recalling the discussion made 
following \eqref{eqn:chiExp}, the actual coefficient found on the 
$(1-e_t^2)^{-13/2}$ term is $+73.6036$, which compares well with the second 
number in \eqref{eqn:chiasympnum}.  The asymptotic analysis has thus again 
provided remarkably sharp estimates for an eccentricity singular factor.
\footnote{Note added in proof: while this paper was in press the authors 
became aware that similar asymptotic analysis of hereditary terms was being 
pursued by N. Loutrel and N. Yunes \cite{LoutYune16}.}

\subsection{Using Darwin eccentricity $e$ to map 
$\mathcal{I}(e_t)$ and $\mathcal{K}(e_t)$ 
to $\tilde{\mathcal{I}}(e)$ and $\tilde{\mathcal{K}}(e)$ 
}
\label{sec:EnhanceDarwin}

Our discussion thus far has given the PN energy flux in terms of the standard 
QK time eccentricity $e_t$ in modified harmonic gauge \cite{ArunETC09a}.  
The motion is only known presently to 3PN relative order, which means that 
the QK representation can only be transformed between gauges up to and 
including $y^3$ corrections.  At the same time, our BHP calculations 
accurately include all relativistic effects that are first order in the mass 
ratio.  It is possible to relate the relativistic (Darwin) eccentricity $e$ to 
the QK $e_t$ (in, say, modified harmonic gauge) up through correction terms 
of order $y^3$,
\begin{align}
\label{eqn:etToe}
\frac{e_t^2}{e^2} = &1 - 6 y 
-\frac{\left(15-19\sqrt{1-e^2}\right) 
+ \left(15\sqrt{1-e^2}-15\right)e^2}{(1-e^2)^{3/2}}y^2 \notag \\
&\quad +\frac{1}{(1-e^2)^{5/2}}
\biggl[
\left(30-38\sqrt{1-e^2}\right)
+\left(59\sqrt{1-e^2}-75\right)e^2
+\left(45-18\sqrt{1-e^2}\right)e^4
\biggr]y^3 .
\end{align}
See \cite{ArunETC09a} for the low-eccentricity limit of this more general 
expression.  We do not presently know how to calculate $e_t$ beyond this 
order.  Using this expression we can at least transform expected fluxes to 
their form in terms of $e$ and check current PN results through 3PN order.  
However, to go from 3PN to 7PN, as we do in this paper, our results must be
given in terms of $e$. 

The instantaneous ($\mathcal{I}$) and hereditary ($\mathcal{K}$) flux terms 
may be rewritten in terms of the relativistic eccentricity $e$ 
straightforwardly by substituting $e$ for $e_t$ using \eqref{eqn:etToe} in 
the full 3PN flux \eqref{eqn:energyflux} and re-expanding the result in powers 
of $y$.  All flux coefficients that are lowest order in $y$ are unaffected 
by this transformation.  Instead, only higher order corrections are 
modified.  We find
\begin{align}
\label{eqn:tiIK0}
\tilde{\mathcal{I}}_0(e) &= \mathcal{I}_0(e_t) , \qquad
\tilde{\mathcal{K}}_{3/2}(e) = \mathcal{K}_{3/2}(e_t) , \qquad
\tilde{\mathcal{K}}_3(e) = \mathcal{K}_3(e_t) , 
\\
\label{eqn:tiI1}
\tilde{\mathcal{I}}_1(e) &= \frac{1}{(1-e^2)^{9/2}}
\left(-\frac{1247}{336}-\frac{15901}{672} e^2-\frac{9253}{384} e^4
-\frac{4037}{1792} e^6
\right),\\
\label{eqn:tiI2}
\tilde{\mathcal{I}}_2(e) &= 
\frac{1}{(1-e^2)^{11/2}}
{\left(-\frac{203471}{9072}-\frac{1430873 }{18144}e^2+\frac{2161337}{24192} e^4
+\frac{231899}{2304} e^6+\frac{499451}{64512} e^8\right)}
\\
&\hspace{40ex} + \frac{1}{(1-e^2)^{5}} \left(\frac{35}{2} 
+ \frac{1715}{48} e^2 
  - \frac{2975}{64} e^4 - \frac{1295}{192} e^6 \right) , \notag \\
\label{eqn:tiI3}
\tilde{\mathcal{I}}_3(e) &= \frac{1}{(1-e^2)^{13/2}}
\left(
\frac{2193295679}{9979200}+\frac{55022404229 }{19958400}e^2
+\frac{68454474929 }{13305600}e^4 
\right.
\\
\biggl.
&\hspace{30ex}+ 
\frac{40029894853}{26611200} e^6
-\frac{32487334699 }{141926400}e^8
-\frac{233745653 }{11354112}e^{10}
\biggr) \notag
\\&
\qquad + \frac{1}{(1-e^2)^{6}} \left(
-\frac{14047483}{151200}-\frac{75546769}{100800}e^2-\frac{210234049}{403200}e^4
+\frac{1128608203}{2419200} e^6
+\frac{617515}{10752} e^8
\right) \notag
\\&
\qquad +\frac{1712}{105} \log\left[\frac{y}{y_0}
  \frac{1+\sqrt{1-e^2}}{2(1-e^2)}\right] F(e) , \notag 
\\
\label{eqn:tiK52}
\tilde{\mathcal{K}}_{5/2}(e) &=\frac{-\pi}{(1-e^2)^6}
\biggl(
\frac{8191}{672}+\frac{62003}{336} e^2
+\frac{20327389}{43008} e^4
+\frac{87458089}{387072} e^6
+\frac{67638841}{7077888} e^8
+\frac{332887 }{25804800}e^{10}
\notag\\&\hspace{20ex}
-\frac{482542621}{475634073600} e^{12}
+\frac{43302428147}{69918208819200} e^{14}
-\frac{2970543742759 }{35798122915430400}e^{16}
\notag\\&\hspace{20ex}+\frac{3024851376397}{207117711153561600} e^{18}
+ \frac{24605201296594481}{4639436729839779840000} e^{20}+\cdots
\biggr) ,
\end{align}
where $F$ is given by \eqref{eqn:capFe} with $e_t \rightarrow e$.  The full 
3PN flux is written exactly as Eqn.~\eqref{eqn:energyflux} with 
$\mathcal{I}\to\tilde{\mathcal{I}}$ and $\mathcal{K}\to\tilde{\mathcal{K}}$.

\section{Confirming eccentric-orbit fluxes through 3PN relative order}
\label{sec:confirmPN}

Sections \ref{sec:homog} and \ref{sec:InhomogSol} briefly described a 
formalism for an efficient, arbitrary-precision MST code for use with 
eccentric orbits.  Section \ref{sec:preparePN} detailed new high-order 
expansions in $e^2$ that we have developed for the hereditary PN terms.  The 
next goal of this paper is to check all known PN coefficients for the energy 
flux (at lowest order in the mass ratio) for eccentric orbits.  The MST code 
is written in \emph{Mathematica} to allow use of its arbitrary precision 
functions.  Like previous circular orbit calculations 
\cite{ShahFrieWhit14,Shah14}, we employ very high accuracy calculations (here 
up to 200 decimal places of accuracy) on orbits with very wide separations 
($p \simeq 10^{15} - 10^{35}$).  Why such wide separations?  At $p = 10^{20}$, 
successive terms in a PN expansion separate by 20 decimal places from each 
other (10 decimal places for half-PN order jumps).  It is like doing QED 
calculations and being able to dial down the fine structure constant from 
$\alpha \simeq 1/137$ to $10^{-20}$.  This in turn mandates the use of 
exceedingly high-accuracy calculations; it is only by calculating with 200 
decimal places that we can observe $\sim 10$ PN orders in our numerical 
results with some accuracy.

\subsection{Generating numerical results with the MST code}
\label{sec:CodeDetails}

In Secs.~\ref{sec:homog} and \ref{sec:InhomogSol} we covered the 
theoretical framework our code uses.  We now provide an algorithmic 
roadmap for the code.  (While the primary focus of this paper is in 
computing fluxes, the code is also capable of calculating local quantities 
to the same high accuracy.)

\begin{itemize}

\item \emph{Solve orbit equations for given $p$ and $e$.}  Given a set of 
orbital parameters, we find $t_p (\chi)$, $\vp_p (\chi)$, and $r_p(\chi)$ to 
high accuracy at locations equally spaced in $\chi$.  We do so by employing 
the SSI method outlined in Sec.~II B of Ref.~\cite{HoppETC15}.  From these 
functions we also obtain the orbital frequencies $\O_r$ and $\O_\vp$.  
All quantities are computed with some pre-determined overall accuracy goal; 
in this paper it was a goal of 200 decimal places of accuracy in the energy 
flux.

\item \emph{Obtain homogeneous solutions to the FD RWZ master equation 
for given $lmn$ mode.}  We find the homogeneous solutions using the MST 
formalism outlined in Sec.~\ref{sec:MST}.  The details of the calculation 
are given here. 

\begin{enumerate}

\item \emph{Solve for $\nu$.}  For each $lmn$, the $\o$-dependent 
renormalized angular momentum $\nu$ is determined (App.~\ref{sec:solveNu}).

\item \emph{Determine at what $n$ to truncate infinite MST sums involving 
$a_n$.}  The solutions $R_{lm\omega}^\text{up/in}$ are infinite sums 
\eqref{eqn:Down1} and \eqref{eqn:RMinus}.  Starting with $a_0=1$, we determine 
$a_n$ for $n<0$ and $n>0$ using Eqn.~\eqref{eqn:RL}.  Terms are added on 
either side of $n=0$ until the homogeneous Teukolsky equation is satisfied 
to within some error criterion at a single point on the orbit.  In the 
post-Newtonian regime the behavior of the size of these terms is well 
understood \cite{ManoSuzuTaka96a,ManoSuzuTaka96b,KavaOtteWard15}.  Our 
algorithm tests empirically for stopping.  Note that 
in addition to forming 
$R_{lm\omega}^\text{up/in}$, residual testing requires computing its first 
and second derivatives.  Having tested for validity of the stopping criterion 
at one point, we spot check at other locations around the orbit.  Once the 
number of terms is established we are able to compute the Teukolsky function 
and its first derivative at any point along the orbit.  (The index $n$ here 
is not to be confused with the harmonic index on such functions as 
$\hat{X}_{lmn}$.)

\item \emph{Evaluate Teukolsky function between $r_{\rm min}$ and 
$r_{\rm max}$.}  Using the truncation of the infinite MST series, we evaluate
$R_{lm\omega}^\text{up/in}$ and their first derivative [higher derivatives 
are found using the homogeneous differential equation \eqref{eqn:radial}] 
at the $r$ locations corresponding to the even-$\chi$ spacing found
in Step 1.  The high precision evaluation of hypergeometric functions 
in this step represents the computational bottleneck in the code.

\item \emph{Transform Teukolsky function to RWZ master functions.}  For 
$l+m$ odd we use Eqn.~\eqref{eqn:RWtrans1} to obtain $\hat{X}_{lmn}^\pm$.  
When $l+m$ is even we continue and use Eqn.~\eqref{eqn:RWtrans2}. 

\item \emph{Scale master functions.}  In solving for the fluxes, it is 
convenient to work with homogeneous solutions that are unit-normalized at
the horizon and at infinity.  We divide the RWZ solutions by the asymptotic 
amplitudes that arise from choosing $a_0 = 1$ when forming the MST solutions
to the Teukolsky equation.  These asymptotic forms are given in
Eqns.~\eqref{eqn:RWasymp1}-\eqref{eqn:RWasymp3}. 

\end{enumerate}
\item \emph{Form $lmn$ flux contribution.}  Form $C_{lmn}^+$ using the 
exponentially-convergent SSI sum \eqref{eqn:CfromEbar}.  Note that this 
exponential convergence relies on the fact that we evaluated the homogeneous 
solutions at evenly-spaced locations in $\chi$.  The coefficient 
$C_{lmn}^+$ feeds into a single positive-definite term in the sum 
\eqref{eqn:fluxNumeric}.

\item \emph{Sum over $lmn$ modes.}  In reconstructing the total flux there 
are three sums:

\begin{enumerate}
\item \emph{Sum over $n$.}  For each spherical harmonic $lm$, there is 
formally an infinite Fourier series sum from $n=-\infty$ to $\infty$.  
In practice the SSI method shows that $n$ is effectively bounded in some range
$-N_1\leq n \leq N_2$.  This range is determined by the fineness of the 
evenly-spaced sampling of the orbit in $\chi$.  For a given orbital sampling, 
we sum modes between $-N_1 \leq n \leq N_2$, where $N_1$ and $N_2$ are the 
first Nyquist-like notches in frequency, beyond which aliasing effects set in 
\cite{HoppETC15}. 

\item \emph{Sum over $m$.}  For each $l$ mode, we sum over $m$ from 
$-l\leq m \leq l$.  In practice, symmetry allows us to sum from $0\leq m
\leq l$, multiplying positive $m$ contributions by 2. 

\item \emph{Sum over $l$.}  The sum over $l$ is, again, formally infinite.  
However, each multipole order appears at a higher PN order, the spacing of 
which depends on $1/p$.  The leading $l=2$ quadrupole flux appears at 
$\mathcal{O}(p^{-5})$.  For an orbit with $p=10^{20}$, the $l=3$ flux appears
at a level 20 orders of magnitude smaller.  Only contributions through 
$l\leq 12$ are necessary with this orbit and an overall precision goal of 
200 digits.  This cutoff in $l$ varies with different $p$.  

\end{enumerate}
\end{itemize}

\subsection{Numerically confirming eccentric-orbit PN results through 
3PN order}
\label{sec:Result1}

We now turn to confirming past eccentric-orbit PN calculations.  The MST 
code takes as input the orbital parameters $p$ and $e$.  Then $1/p$ 
is a small parameter.  Expanding $dt_p/d\chi$ in \eqref{eqn:darwinEqns} we 
find from \eqref{eqn:O_r} 
\begin{align}
\label{eqn:O_rPN}
\O_r = \frac{1}{M} \l \frac{1-e^2}{p} \r^{3/2} 
\left\{ 1 
- 3 \frac{1-e^2}{p} 
- \frac{3}{2} \frac{\sqrt{1 - e^2} 
[5 - 2 \sqrt{1 - e^2} + e^2 (-5 + 6 \sqrt{1 - e^2} )]}{p^2}  
+ \cdots
\right\} .
\end{align}
Expanding \eqref{eqn:O_phi} in similar fashion gives
\begin{align}
\label{eqn:O_phiPN}
\O_\vp = \frac{1}{M} \l \frac{1-e^2}{p} \r^{3/2} 
\left\{ 1 
+ 3 \frac{e^2}{p} 
- \frac{3}{4} \frac{\left[10 (-1 + \sqrt{1 - e^2}) + 
e^2 (20 - 3 \sqrt{1 - e^2}) + 2 e^4 (-5 + 6 \sqrt{1 - e^2}) \right]}
{\sqrt{1 - e^2} p^2} 
+ \cdots
\right\}.
\end{align}
Then given the definition of $y$ we obtain an expansion of $y$ in terms 
of $p$
\begin{align}
y = \frac{1 - e^2}{p} 
+ \frac{2 e^2 (1 - e^2)}{p^2} 
+ \frac{1}{2} \sqrt{1 - e^2} \frac{\left[10 (1 - \sqrt{1 - e^2}) 
+ e^2 (-3 + 10 \sqrt{1 - e^2}) + 10 e^4) \right]}{p^3}
+ \cdots .
\end{align}
So from our chosen parameters $e$ and $p$ we can obtain $y$ to arbitrary 
accuracy, and then other orbital parameters, such as $\Omega_r$ and 
$\Omega_\vp$, can be computed as well to any desired accuracy.

\begin{figure}
\centering
\begin{minipage}{0.48\textwidth}
\centering
\includegraphics[scale=0.98]{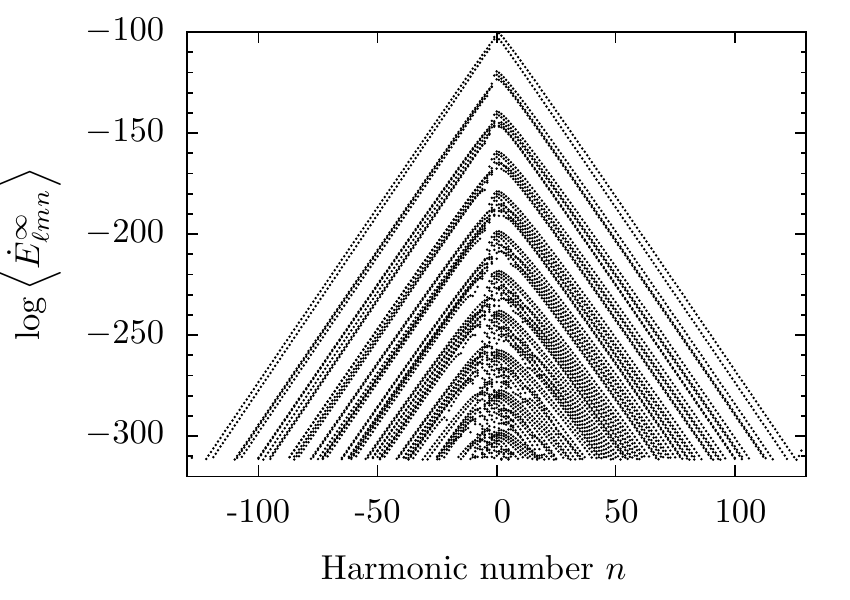}
\caption{Fourier-harmonic energy-flux spectra from an orbit with semi-latus 
rectum $p=10^{20}$ and eccentricity $e=0.1$.  Each inverted-$V$ spectrum 
represents flux contributions of modes with various harmonic number $n$ but 
fixed $l$ and $m$.  The tallest spectrum traces the harmonics of the $l=2$, 
$m=2$ quadrupole mode, the dominant contributor to the flux.  Spectra of 
successively higher multipoles (octupole, hexadecapole, etc) each drop 20 
orders of magnitude in strength as $l$ increases by one ($l \leq 12$ are 
shown).  Every flux contribution is computed that is within 200 decimal 
places of the peak of the quadrupole spectrum.  With $e = 0.1$, there were 
7,418 significant modes that had to be computed (and are shown above).
\label{fig:fluxSpectra}} 
\end{minipage}\hfill
\begin{minipage}{0.48\textwidth}
\centering
\includegraphics[scale=0.98]{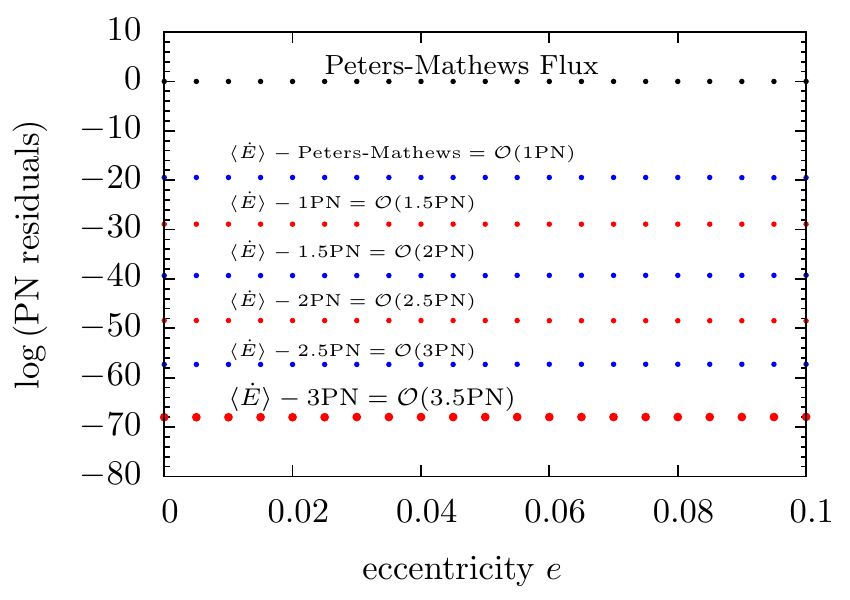}
\caption{Residuals after subtracting from the numerical data successive PN 
contributions.  Residuals are shown for a set of orbits with $p = 10^{20}$ 
and a range of eccentricities from $e = 0.005$ through $e = 0.1$ in steps 
of $0.005$.  Residuals are scaled relative to the Peters-Mathews 
flux (uppermost points at unit level).  The next set of points (blue) shows 
residuals after subtracting the Peters-Mathews enhancement from BHP data.  
Residuals drop uniformly by 20 order of magnitude, consistent with 1PN 
corrections in the data.  The next (red) points result from subtracting the 
1PN term, giving residuals at the 1.5PN level.  Successive subtraction of 
known PN terms is made, reaching final residuals at 70 orders of magnitude 
below the total flux and indicating the presence of 3.5PN contributions in 
the numerical fluxes.  
\label{fig:residuals1}} 
\end{minipage}
\end{figure}

To check past work \cite{ArunETC08a,ArunETC08b,ArunETC09a,Blan14} on the PN 
expansion of the energy flux, we used a single orbital separation 
($p = 10^{20}$), with a set of eccentricities ($e = 0.005$ through $e=0.1$).
For each $e$, we compute the flux for each $lmn$-mode up 
to $l = 12$ to as much as 200 decimal places of accuracy (the accuracy can 
be relaxed for higher $l$ as these modes contribute only weakly to the total 
energy flux).  Fig.~\ref{fig:fluxSpectra} depicts all 
7,418 $lmn$-modes that contributed to the energy flux for just the $p = 10^{20}$, 
$e = 0.1$ orbit.  Making this calculation and sum for all $e$, we then have 
an eccentricity dependent flux.  Next, we compute the PN parts of the 
expected flux using Eqns.~\eqref{eqn:tiIK0} through \eqref{eqn:tiK52}.  The 
predicted flux $\mathcal{F}_{\rm 3PN}$ is very close to the computed flux 
$\mathcal{F}_{\rm MST}$.  We then subtract the quadrupole theoretical flux 
term 
\be
\mathcal{F}_{\rm N} = \frac{32}{5} \left(\frac{\mu}{M}\right)^2 y^5 \, 
\tilde{\mathcal{I}}_0(e) 
\equiv 
\mathcal{F}^{\rm circ}_{\rm N} \ \tilde{\mathcal{I}}_0(e) ,
\ee
from the flux computed with the MST code (and normalize with respect to the 
Newtonian term) 
\be
\frac{\mathcal{F}_{\rm MST} - \mathcal{F}_{\rm N}}{\mathcal{F}_{\rm N}} 
= \mathcal{O}(y) 
\simeq \frac{1}{\tilde{\mathcal{I}}_0 (e)} \left[y\,\tilde{\mathcal{I}}_1(e) 
+ y^{3/2}\,\tilde{\mathcal{K}}_{3/2}(e) 
+ y^2\,\tilde{\mathcal{I}}_2(e) 
+ y^{5/2}\,\tilde{\mathcal{K}}_{5/2}(e) 
+ y^3\,\tilde{\mathcal{I}}_3(e) 
+ y^3\,\tilde{\mathcal{K}}_{3}(e) \right] ,
\ee
and find a residual that is 20 orders of magnitude smaller than the 
quadrupole flux.  The residual reflects the fact that our numerical (MST 
code) data contains the 1PN and higher corrections.  We next subtract the 
1PN term
\be
\frac{\mathcal{F}_{\rm MST} - \mathcal{F}_{\rm N}}{\mathcal{F}_{\rm N}} 
- y\, \frac{\tilde{\mathcal{I}}_1(e) }{\tilde{\mathcal{I}}_0(e)}
= \mathcal{O}(y^{3/2}) ,
\ee
and find residuals that are another 10 orders of magnitude lower.  This 
reflects the expected 1.5PN tail correction.  Using our high-order expansion 
for $\vp(e)$, we subtract and reach 2PN residuals.  Continuing in this way, 
once the 3PN term is subtracted, the residuals lie at a level 70 orders of 
magnitude below the quadrupole flux.  We have reached the 3.5PN 
contributions, which are encoded in the MST result but whose form is 
(heretofore) unknown.  Fig.~\ref{fig:residuals1} shows this process 
of successive subtraction.  \emph{We conclude that the published PN 
coefficients} \cite{ArunETC08a,Blan14} \emph{for eccentric orbits in the lowest 
order in $\nu$ limit are all correct.}  Any error would have to be at the 
level of one part in $10^{10}$ (and only then in the 3PN terms) or it would 
show up in the residuals.

As a check we made this comparison also for other orbital radii and using the 
original expressions in terms of $e_t$ (which we computed from $e$ and $y$ 
to high precision).  The 2008 results \cite{ArunETC08a} continued to stand.

\section{Determining new PN terms between orders 3.5PN and 7PN}
\label{sec:newPN}

Having confirmed theoretical results through 3PN, we next sought to determine 
analytic or numerical coefficients for as-yet unknown PN coefficients at 
3.5PN and higher orders.  We find new results to 7PN order.  

\subsection{A model for the higher-order energy flux}

The process begins with writing an expected form for the expansion.  As 
discussed previously, beyond 3PN we presently do not know $e_t$, so all new 
results are parameterized in terms of the relativistic $e$ (and $y$).  Based 
on experience with the expansion up to 3PN (and our expansions of the 
hereditary terms), we build in the expected eccentricity singular factors 
from the outset.  In addition, with no guidance from analytic PN theory, we 
have no way of separating instantaneous from hereditary terms beyond 3PN 
order, and thus denote all higher-order PN enhancement factors with the 
notation $\mathcal{L}_i(e)$.  Finally, known higher-order work \cite{Fuji12} 
in the circular-orbit limit allows us to anticipate the presence of various 
logarithmic terms and powers of logs.  Accordingly, we take the flux to 
have the form 
\begin{align}
\label{eqn:energyfluxNew}
\left\langle \frac{dE}{dt} \right\rangle =  
&\mathcal{F}_{\rm 3PN} 
+\mathcal{F}^{\rm circ}_{\rm N} 
\biggl[\mathcal{L}_{7/2}y^{7/2}
+y^4\Bigl(\mathcal{L}_4+\log(y)\mathcal{L}_{4L}\Bigr)
+y^{9/2}\Bigl(\mathcal{L}_{9/2}+\log(y)\mathcal{L}_{9/2L}\Bigr)
+y^5\Bigl(\mathcal{L}_5
+\log(y)\mathcal{L}_{5L}\Bigr)
\notag\\&\quad 
+y^{11/2}\Bigl(\mathcal{L}_{11/2}+\log(y)\mathcal{L}_{11/2L}\Bigr)
+ y^6\Bigl(\mathcal{L}_6 + \log(y)\mathcal{L}_{6L}
+ \log^2(y)\mathcal{L}_{6L^2} \Bigr)
+y^{13/2}\Bigl(\mathcal{L}_{13/2}+\log(y)\mathcal{L}_{13/2L}\Bigr)
\notag\\&\quad 
+ y^7\Bigl(\mathcal{L}_7 + \log(y)\mathcal{L}_{7L}
+ \log^2(y)\mathcal{L}_{7L^2} \Bigr)
+ y^{15/2}\Bigl(\mathcal{L}_{15/2} + \log(y)\mathcal{L}_{15/2L}
+ \log^2(y)\mathcal{L}_{15/2L^2}\Bigr)
\notag\\&\quad 
+ y^{8}\Bigl(\mathcal{L}_{8} + \log(y)\mathcal{L}_{8L}
+ \log^2(y)\mathcal{L}_{8L^2}\Bigr)
+ y^{17/2}\Bigl(\mathcal{L}_{17/2} + \log(y)\mathcal{L}_{17/2L}
+ \log^2(y)\mathcal{L}_{17/2L^2}\Bigr)
\notag\\&\quad 
+ y^{9}\Bigl(\mathcal{L}_{9} + \log(y)\mathcal{L}_{9L}
+ \log^2(y)\mathcal{L}_{9L^2}
+ \log^3(y)\mathcal{L}_{9L^3}\Bigr)
\notag\\&\quad 
+ y^{19/2}\Bigl(\mathcal{L}_{19/2} + \log(y)\mathcal{L}_{19/2L}
+ \log^2(y)\mathcal{L}_{19/2L^2}
+ \log^3(y)\mathcal{L}_{19/2L^3}\Bigr)
\notag\\&\quad +
y^{10}\Bigl(\mathcal{L}_{10} + \log(y)\mathcal{L}_{10L}
+ \log^2(y)\mathcal{L}_{10L^2}
+ \log^3(y)\mathcal{L}_{10L^3}\Bigr)
\biggr].
\end{align}
It proves useful to fit MST code data all the way through 10PN order even 
though we quote new results only up to 7PN.  

\subsection{Roadmap for fitting the higher-order PN expansion}

The steps in making a sequence of fits to determine the higher-order PN 
expansion are as follows:

\begin{itemize}
\item \emph{Compute data for orbits with various $e$ and $y$}.  We compute 
fluxes for 1,683 unique orbits, with 33 eccentricities for each of 51 
different orbital separations ($p$ or $y$ values).  The models include 
circular orbits and eccentricities 
ranging from $e = 10^{-5}$ to $e = 0.1$.  The $p$ range is from $10^{10}$ 
through $10^{35}$ in half-logarithmic steps, i.e., $10^{10},10^{10.5},\dots$.
The values of $y$ are derived from $p$ and $e$.

\item \emph{Use the expected form of the expansion in $y$}.  As mentioned 
earlier, known results for circular fluxes on Schwarzschild backgrounds 
allow us to surmise the expected terms in the $y$-expansion, shown in 
Eqn.~\eqref{eqn:energyfluxNew}.  The expansion through 10PN order contains 
as a function of $y$ 44 parameters, which can be determined by our dataset 
with 51 $y$ values (at each eccentricity).

\item \emph{Eliminate known fit parameters}.  The coefficients at 0PN, 
1PN, 1.5PN, 2PN, and 3PN relative orders involve known enhancement functions 
of the eccentricity $e$ (given in the previous section) and these terms may 
be subtracted from the data and excluded from the fit model.  It is important 
to note that we do not include the 2.5PN term in this subtraction.  Though 
we have a procedure for expanding the $\mathcal{K}_{5/2}$ term to high order 
in $e^2$, it has proven computationally difficult so far to expand beyond 
$e^{70}$.  This order was sufficient for use in Sec.~\ref{sec:confirmPN} in 
confirming prior results to 3PN but is not accurate enough to reach 10PN 
(at the large radii we use).  We instead include a parameterization of 
$\mathcal{K}_{5/2}$ in the fitting model.

\item \emph{Fit for the coefficients on powers of $y$ and $\log(y)$}.  
We use \emph{Mathematica's} \texttt{NonlinearModelFit} function to obtain 
numerical values for the coefficients $\mathcal{L}_{7/2}$, $\mathcal{L}_4$, 
$\dots$ shown in Eqn.~\eqref{eqn:energyfluxNew}.  We perform this fit 
separately for each of the 33 values of $e$ in the dataset.  

\item \emph{Organize the numerically determined functions of $e$ for each 
separate coefficient $\mathcal{L}_i(e)$ in the expansion over $y$ and 
$\log(y)$}.  Having fit to an expansion of the form \eqref{eqn:energyfluxNew} 
and eliminated known terms there remain 38 functions of $e$, each of which 
is a discrete function of 33 different eccentricities.

\item \emph{Assemble an expected form for the expansion in $e$ of 
each $\mathcal{L}_i(e)$}.  Based on the pattern in Sec.~\ref{sec:preparePN}, 
each full (or half) PN order $= N$ will have a leading eccentricity 
singular factor of the form $(1-e^2)^{-7/2 - N}$.  The remaining power 
series will be an expansion in powers of $e^2$.

\item \emph{Fit each model for $\mathcal{L}_i(e)$ using data ranging over 
eccentricity}.  The function \texttt{NonlinearModelFit} is again used to 
find the unknown coefficients in the eccentricity function expansions.  
With data on 33 eccentricities, the coefficient models are limited to 
at most 33 terms.  However, it is possible to do hierarchical fitting.  As 
lower order coefficients are firmly determined in analytic form (see next 
step), they can be eliminated in the fitting model to allow new, higher-order 
ones to be included.
 
\item \emph{Attempt to determine analytic form of $e^2$ coefficients}.
It is possible in some cases to determine the exact analytic form (rational 
or irrational) of coefficients of $e^2$ determined only in decimal value 
form in the previous step.  We use \emph{Mathematica}'s function 
\texttt{FindIntegerNullVector} (hereafter FINV), which is an implementation 
of the PSLQ integer-relation algorithm. 

\item \emph{Assess the validity of the analytic coefficients}.
A rational or irrational number, or combination thereof, predicted by FINV 
to represent a given decimal number has a certain probability of being a 
coincidence (note: the output of FINV will still be a very accurate 
\emph{representation} of the input decimal number).  If FINV outputs, say, 
a single rational number with $N_N$ digits in its numerator and $N_D$ digits 
in its denominator, and this rational agrees with the input decimal number 
it purports to represent to $N$ digits, then the likelihood that this is a 
coincidence is of order $\mathcal{P} \simeq 10^{N_N+N_D-N}$ 
\cite{ShahFrieWhit14}.  With the analytic coefficients we give in what 
follows, in no case is the probability of coincidence larger than $10^{-6}$, 
and in many cases the probability is as low as $10^{-90} - 10^{-50}$.  Other 
consistency checks are made as well.  It is important that the analytic 
output of PSLQ not change when the number of significant digits in the input 
is varied (within some range).  Furthermore, as we expect rational numbers 
in a perturbation expansion to be sums of simpler rationals, a useful 
criterion for validity of an experimentally determined rational is that it 
have no large prime factors in its denominator \cite{JohnMcDaShahWhit15}.

\end{itemize}

\subsection{The energy flux through 7PN order}

We now give, in mixed analytic and numeric form, the PN expansion (at lowest 
order in $\nu$) for the energy flux through 7PN order.  Analytic coefficients 
are given directly, while well-determined coefficients that are known only 
in numerical form are listed in the formulae as numbered parameters like 
$b_{26}$.  The numerical values of these coefficients are tabulated in 
App.~\ref{sec:numericEnh}.  We find for the 3.5PN and 4PN (non-log) 
terms
\begin{align}
\mathcal{L}_{7/2} &= -\frac{\pi}{(1-e^2)^7}
\biggl(\frac{16285}{504}+ \frac{21500207}{48384}e^2 +  
\frac{3345329}{48384}e^4 -  \frac{111594754909}{41803776} e^6-  
\frac{82936785623}{55738368} e^8 - \frac{11764982139179}{107017666560} e^{10} 
\notag\\&\ -
\frac{216868426237103}{9631589990400} e^{12} 
- \frac{30182578123501193}{2517055517491200}e^{14}  
-  \frac{351410391437739607}{48327465935831040} e^{16}
-  \frac{1006563319333377521717}{208774652842790092800}e^{18} 
\notag\\&\ -
\frac{138433556497603036591}{40776299383357440000}e^{20}
 - \frac{16836217054749609972406421}{6736462131727360327680000}e^{22}
-  \frac{2077866815397007172515220959}{1091306865339832373084160000}e^{24}
\notag\\&\ +
b_{26}e^{26}+ b_{28}e^{28}+ b_{30}e^{30}+ b_{32}e^{32}
+ b_{34}e^{34}+ b_{36}e^{36}+ b_{38}e^{38}+ b_{40}e^{40}
+ b_{42}e^{42}+ b_{44}e^{44}+ b_{46}e^{46}+ b_{48}e^{48}
\notag\\&\ +
b_{50}e^{50}+ b_{52}e^{52}+ b_{54}e^{54}
+\cdots
\biggr),\label{eqn:Ienh72}\\
\notag \\
\mathcal{L}_4 &=\frac{1}{(1-e^2)^{15/2}}
\biggl[
-\frac{323105549467}{3178375200}+\frac{232597}{4410}\gamma_\text{E}
-\frac{1369}{126}\pi^2+\frac{39931}{294}\log(2)
-\frac{47385}{1568}\log(3)
\notag\\&\ +
\biggl(-\frac{128412398137}{23543520}
+\frac{4923511}{2940}\gamma_\text{E}
-\frac{104549}{252}\pi^2-\frac{343177}{252} \log(2) +\frac{55105839 }{15680}\log(3)
\biggr)e^2
\notag\\&\ 
+\biggl(-\frac{981480754818517}{25427001600}
+\frac{142278179}{17640}\gamma_\text{E}-\frac{1113487}{504}\pi^2
+\frac{762077713}{5880}\log(2)
-\frac{2595297591}{71680}\log(3)
\notag\\&\hspace{25ex}-\frac{15869140625}{903168}\log(5)\biggr)e^4
\notag\\&\ +\biggl(-\frac{874590390287699}{12713500800}
+\frac{318425291}{35280}\gamma_\text{E}
-\frac{881501}{336}\pi^2-\frac{90762985321}{63504}\log(2)
+\frac{31649037093}{1003520}\log(3)
\notag\\&\hspace{25ex}+\frac{10089048828125}{16257024}\log(5)\biggr)e^6 
\notag\\&\ + d_{8}e^{8} + d_{10}e^{10}
 + d_{12}e^{12} + d_{14}e^{14} + d_{16}e^{16} + d_{18}e^{18}
  + d_{20}e^{20} + d_{22}e^{22} + d_{24}e^{24} + d_{26}e^{26}
  \notag\\&\ + d_{28}e^{28}
   + d_{30}e^{30} + d_{32}e^{32} + d_{34}e^{34} + d_{36}e^{36} + d_{38}e^{38}
   + d_{40}e^{40}+\cdots
\biggr].\label{eqn:Ienh4}
\end{align}
In both of these expressions the circular orbit limits ($e^0$) were known 
\cite{Fuji12}.  These results have been presented earlier \cite{Fors14,Fors15,
Evan15a} and are available online.  The coefficients through $e^6$ for 3.5PN 
and 4PN are also discussed in \cite{SagoFuji15}, which we found to be in 
agreement with our results.  We next consider the 4PN log contribution, which 
we find to have an \emph{exact, closed-form expression}
\begin{align}
\mathcal{L}_{4L} &= \frac{1}{(1-e^2)^{15/2}}
\biggl(
\frac{232597}{8820} + \frac{4923511}{5880}e^2
+ \frac{142278179}{35280}e^4+
\frac{318425291}{70560}e^6
+ \frac{1256401651}{1128960}e^8
+\frac{7220691}{250880}e^{10}
\biggr) . \label{eqn:Ienh4L}
\end{align}
In the 4.5PN non-log term we were only able to find analytic coefficients for 
the circular limit (known previously) and the $e^2$ term.  We find many 
higher-order terms numerically (App.~\ref{sec:numericEnh})
\begin{align}
\mathcal{L}_{9/2} &= \frac{\pi}{(1-e^2)^8}
\biggl[\frac{265978667519}{745113600}-\frac{6848}{105}\gamma_\text{E}
-\frac{13696}{105}\log(2)
+ \biggl(\frac{5031659060513}{447068160}-\frac{418477 }{252} \gamma_\text{E}
-\frac{1024097 }{1260}\log(2)
\notag\\&\ -\frac{702027}{280}\log(3)\biggr)e^2
+ h_{4}e^{4}+ h_{6}e^{6}+ h_{8}e^{8}+ h_{10}e^{10}
+ h_{12}e^{12}+ h_{14}e^{14}+ h_{16}e^{16}+ h_{18}e^{18}+ h_{20}e^{20}
+ h_{22}e^{22}
\notag\\&\ + h_{24}e^{24}+ h_{26}e^{26}+ h_{28}e^{28}+ h_{30}e^{30}
+ h_{32}e^{32}+ h_{34}e^{34}+ h_{36}e^{36}
\biggr] . \label{eqn:Ienh92}
\end{align}
In the 4.5PN log term we are able to find the first 10 coefficients in 
analytic form and 6 more in accurate numerical form 
(App.~\ref{sec:numericEnh})
\begin{align}
\mathcal{L}_{9/2L} &= -\frac{\pi}{(1-e^2)^8}
\biggl(
\frac{3424}{105} + \frac{418477}{504}e^2+
\frac{32490229}{10080}e^4 + 
\frac{283848209}{96768}e^6+
\frac{1378010735}{2322432}e^8
+\frac{59600244089}{4644864000}e^{10}
\notag\\&\ -
\frac{482765917}{7962624000}e^{12}
+\frac{532101153539}{29132587008000}e^{14}
-\frac{576726373021}{199766310912000}e^{16}
+\frac{98932878601597}{3624559945187328000}e^{18}
+g_{20}e^{20}
\notag\\&\ +g_{22}e^{22}+g_{24}e^{24}+g_{26}e^{26}+g_{28}e^{28}+g_{30}e^{30}+
\cdots
\biggr) . \label{eqn:Ienh92L}
\end{align}
For the 5PN non-log term, we are only able to confirm the circular-orbit 
limit analytically.  Many other terms were found with accurate numerical 
values (App.~\ref{sec:numericEnh})
\begin{align}
\mathcal{L}_5 &= \frac{1}{(1-e^2)^{17/2}}
\biggl[
-\frac{2500861660823683}{2831932303200}
-\frac{424223}{6804} \pi ^2
+\frac{916628467}{7858620} \gamma_\text{E} 
-\frac{83217611 }{1122660}\log(2)+\frac{47385}{196}\log(3)
\notag\\&\ + k_{2}e^{2} + k_{4}e^{4} + k_{6}e^{6}
 + k_{8}e^{8} + k_{10}e^{10} + k_{12}e^{12} + k_{14}e^{14}
  + k_{16}e^{16} + k_{18}e^{18} + k_{20}e^{20} + k_{22}e^{22}
   + k_{24}e^{24}
\notag\\&\ + k_{26}e^{26}+\cdots
\biggr] . \label{eqn:Ienh5}
\end{align}
In the 5PN log term we found the first 13 terms in analytic form, and 
several more numerically (App.~\ref{sec:numericEnh})
\begin{align}
\mathcal{L}_{5L} &= \frac{1}{(1-e^2)^{17/2}}\biggl(
\frac{916628467}{15717240} + \frac{11627266729}{31434480}e^2-
\frac{84010607399}{10478160}e^4-
\frac{67781855563}{1632960}e^6-
\frac{87324451928671}{2011806720}e^8
\notag\\&\ -\frac{301503186907}{29804544}e^{10}-
\frac{752883727}{1290240}e^{12}-
\frac{22176713}{129024}e^{14}-
\frac{198577769}{2064384}e^{16}-
\frac{250595605}{4128768}e^{18}
\notag\\&\ -
\frac{195002899}{4718592}e^{20}-
\frac{280151573}{9437184}e^{22}-
\frac{1675599991}{75497472}e^{24}
+j_{26}e^{26}+j_{28}e^{28}+\cdots
\biggr) . \label{eqn:Ienh5L}
\end{align}
In the 5.5PN non-log term we found analytic forms for the first two terms 
with 8 more in numerical form (App.~\ref{sec:numericEnh})
\begin{align}
\mathcal{L}_{11/2} = \frac{\pi}{(1-e^2)^9}
& \biggl[
\frac{8399309750401}{101708006400}+
\frac{177293}{1176}\gamma_\text{E} 
+\frac{8521283}{17640}\log(2)
-\frac{142155}{784}\log(3)
\notag\\&\ +
\biggl(-\frac{6454125584294467}{203416012800}
+\frac{197515529}{17640} \gamma_\text{E} 
-\frac{195924727}{17640} \log (2)
+\frac{1909251}{80} \log (3)\biggr)e^2
\notag\\&\ 
+ m_{4}e^{4} + m_{6}e^{6} + m_{8}e^{8} + m_{10}e^{10} + m_{12}e^{12}
+ m_{14}e^{14} + m_{16}e^{16} + m_{18}e^{18}
+\cdots
\biggr] . \label{eqn:Ienh112}
\end{align}
The 5.5PN log term yielded analytic forms for the first six terms with 
several more known only numerically (App.~\ref{sec:numericEnh})
\begin{align}
\mathcal{L}_{11/2L} &= \frac{\pi}{(1-e^2)^9}
\biggl(
\frac{177293}{2352} + \frac{197515529}{35280}e^2
+\frac{22177125281}{451584}e^4+
\frac{362637121649}{3386880}e^6+
\frac{175129893794507}{2601123840}e^8
\notag\\&\ +
\frac{137611940506079}{13005619200}e^{10}
+l_{12}e^{12}+l_{14}e^{14}+l_{16}e^{16}+\cdots
\biggr) . \label{eqn:Ienh112L}
\end{align}
We only extracted the circular-orbit limit analytically for the 6PN non-log 
term.  Six more coefficients are known numerically (App.~\ref{sec:numericEnh})
\begin{align}
\mathcal{L}_6 &= \frac{1}{(1-e^2)^{19/2}}
\biggl(\frac{3803225263}{10478160}\pi^2
-\frac{27392}{105}\zeta (3)+\frac{1465472}{11025} \gamma_{\text{E}}^2
-\frac{256}{45}\pi^4-\frac{27392}{315} \gamma_\text{E}  \pi ^2
-\frac{246137536815857}{157329572400} \gamma_\text{E}
\notag\\&\ +\frac{2067586193789233570693}{602387400044430000}
+\frac{5861888}{11025} \log ^2(2)
-\frac{54784}{315} \pi ^2 \log (2)
-\frac{271272899815409}{157329572400} \log (2)
\notag\\&\ +\frac{5861888}{11025} \gamma_\text{E} \log (2)
-\frac{37744140625}{260941824} \log (5)
-\frac{437114506833}{789268480} \log (3)
+n_2e^2 + n_4e^4+n_6e^6+n_8e^8+n_{10}e^{10}
\notag\\&\ +n_{12}e^{12}+\cdots
\biggr) . \label{eqn:Ienh6}
\end{align}
The 6PN log term yielded analytic forms for the first two terms, with 5 
more in numerical form (App.~\ref{sec:numericEnh})
\begin{align}
\mathcal{L}_{6L} &= \frac{1}{(1-e^2)^{19/2}}
\biggl[
-\frac{246137536815857}{314659144800}-\frac{13696}{315}\pi^2
+\frac{1465472}{11025} \gamma_\text{E} 
+\frac{2930944}{11025} \log (2)
\notag\\&\ 
+\biggl(-\frac{25915820507512391}{629318289600} 
- \frac{1773953}{945}\pi^2 
+ \frac{189812971}{33075}\gamma_\text{E} 
+ \frac{18009277}{4725}\log(2) 
+ \frac{75116889}{9800}\log(3)\biggr)e^2
\notag\\&\ 
+p_4e^4+p_6e^6+p_8e^8+p_{10}e^{10}+p_{12}e^{12}+\cdots
\biggr] . \label{eqn:Ienh6L}
\end{align}
The 6PN squared-log term (first instance of such a term) yielded the first 
seven coefficients in analytic form
\begin{align}
\mathcal{L}_{6L^2} &= \frac{1}{(1-e^2)^{19/2}}\biggl(
\frac{366368}{11025} + \frac{189812971}{132300}e^2
+\frac{1052380631}{105840}e^4
+\frac{9707068997}{529200}e^6
+\frac{8409851501}{846720}e^8
+\frac{4574665481}{3386880}e^{10}
\notag\\&\ +\frac{6308399}{301056}e^{12}+\cdots
\biggr) . \label{eqn:Ienh6L2}
\end{align}
At 6.5PN order, we were only able to confirm the circular-orbit limit in 
the non-log term.  Additional terms are known accurately numerically 
(App.~\ref{sec:numericEnh})
\begin{align}
\mathcal{L}_{13/2}&=\frac{\pi}{(1-e^2)^{10}}
\biggl(-\frac{81605095538444363}{20138185267200}
+\frac{300277177}{436590} \gamma_\text{E} 
-\frac{42817273}{71442} \log (2)
+\frac{142155}{98} \log (3)
+r_2e^2+r_4e^4
\notag\\&\ +r_6e^6+r_8e^8+r_{10}e^{10}
+r_{12}e^{12}+\cdots\biggr) . \label{eqn:Ienh132}
\end{align}
In the 6.5PN log term we found the first two coefficients analytically.  
Others are known numerically (App.~\ref{sec:numericEnh})
\begin{align}
\mathcal{L}_{13/2L} &= \frac{\pi}{(1-e^2)^{10}}\biggl(
\frac{300277177}{873180}+\frac{99375022631}{27941760}e^2+s_4e^4
+s_6e^6+s_8e^8+s_{10}e^{10}
+s_{12}e^{12}+\cdots
\biggr) . \label{eqn:Ienh132L}
\end{align}
At 7PN order in the non-log term, we only confirm the leading term.  
Three more terms are known numerically (App.~\ref{sec:numericEnh})
\begin{align}
\mathcal{L}_7 &= \frac{1}{(1-e^2)^{21/2}}
\biggl(\frac{58327313257446476199371189}{8332222517414555760000}
+\frac{531077}{2205} \zeta (3)
+\frac{2621359845833}{2383781400} \pi^2
+\frac{531077}{6615}\gamma_\text{E} \pi^2
-\frac{9523 }{945}\pi^4
\notag\\
&\ 
-\frac{52525903}{154350} \gamma_\text{E}^2
+\frac{9640384387033067}{17896238860500} \gamma_\text{E} 
+\frac{1848015}{5488}\log^2(3)
-\frac{5811697}{2450} \log^2(2)
+\frac{128223}{245} \pi^2 \log(2)
\notag\\&\ 
+\frac{19402232550751339 }{17896238860500}\log(2)
-\frac{142155}{392} \pi^2\log(3)
+\frac{1848015}{2744} \log(2)\log(3)
+\frac{1848015}{2744} \gamma_\text{E}\log (3)
\label{eqn:Ienh7}\\
&\ 
+\frac{9926708984375}{5088365568} \log (5)
-\frac{471188717 }{231525}\gamma  \log (2)
-\frac{6136997968378863 }{1256910054400}\log (3) 
+ t_2e^2+t_4e^4+t_6e^6+\cdots
\biggr) . \notag 
\end{align}
At 7PN order in the log term we found the first two coefficients analytically.  
Three more orders in $e^2$ are known numerically (App.~\ref{sec:numericEnh})
\begin{align}
\mathcal{L}_{7L} &= 
\frac{1}{(1-e^2)^{21/2}}
\biggl[
\frac{9640384387033067}{35792477721000}
+\frac{531077}{13230} \pi ^2
-\frac{52525903 }{154350}\gamma_\text{E}
-\frac{471188717}{463050} \log (2)
+\frac{1848015}{5488} \log (3)
\notag\\&\ 
+\biggl(\frac{5361621824744487121}{28633982176800} 
+ \frac{20170061}{1764}\pi^2 
- \frac{8436767071}{185220}\gamma_\text{E} 
+ \frac{8661528101}{926100}\log(2) 
- \frac{21008472903}{274400}\log(3)\biggr)e^2
\notag\\&\ +u_4e^4+u_6e^6+u_8e^8+\cdots
\biggr] . \label{eqn:Ienh7L}
\end{align}

\begin{center}
\begin{figure*}
\includegraphics[scale=1.35]{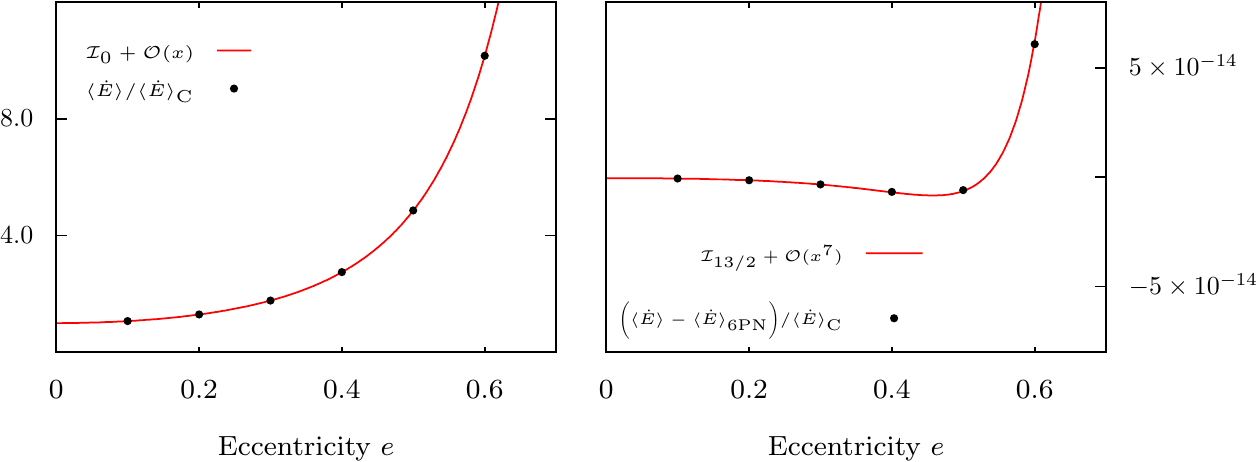}
\caption{Agreement between numerical flux data and the 7PN expansion at 
smaller radius and larger eccentricities.  An orbit with separation of 
$p=10^3$ was used.  The left panel shows the energy flux as a function of 
eccentricity normalized to the circular-orbit limit (i.e., the curve closely 
resembles the Peters-Mathews enhancement function).  The red curve shows the 
7PN fit to this data.  On the right, we subtract the fit (through 6PN order) 
from the energy flux data points.  The residuals have dropped by 14 orders 
of magnitude.  The residuals are then shown to still be well fit by the 
remaining 6.5PN and 7PN parts of the model even for $e=0.6$ .  
\label{fig:twoPanelPlot}} 
\end{figure*}
\end{center}

Finally, at 7PN order there is a squared-log term and we again found the 
first two coefficients analytically
\begin{align}
\mathcal{L}_{7L^2} &= -\frac{1}{(1-e^2)^{21/2}}
\biggl(
\frac{52525903}{617400}+\frac{8436767071}{740880}e^2+v_4e^4+v_6e^6
+v_8e^8+\cdots
\biggr) \qquad \qquad \qquad . \label{eqn:Ienh7L2}
\end{align}

\end{widetext}

\subsection{Discussion}

The analytic forms for the $e^2$ coefficients at the 5.5PN non-log, 6PN log, 
6.5PN log, 7PN log, and 7PN log-squared orders were previously obtained by 
Johnson-McDaniel \cite{JohnMcDa15}.  They are obtained by using the eccentric 
analogue of the simplification described in \cite{JohnMcDa14} to predict 
leading logarithmic-type terms to all orders, starting from the expressions 
for the modes given in Appendix G of \cite{MinoETC97}.  

\begin{figure}
\includegraphics[scale=0.67]{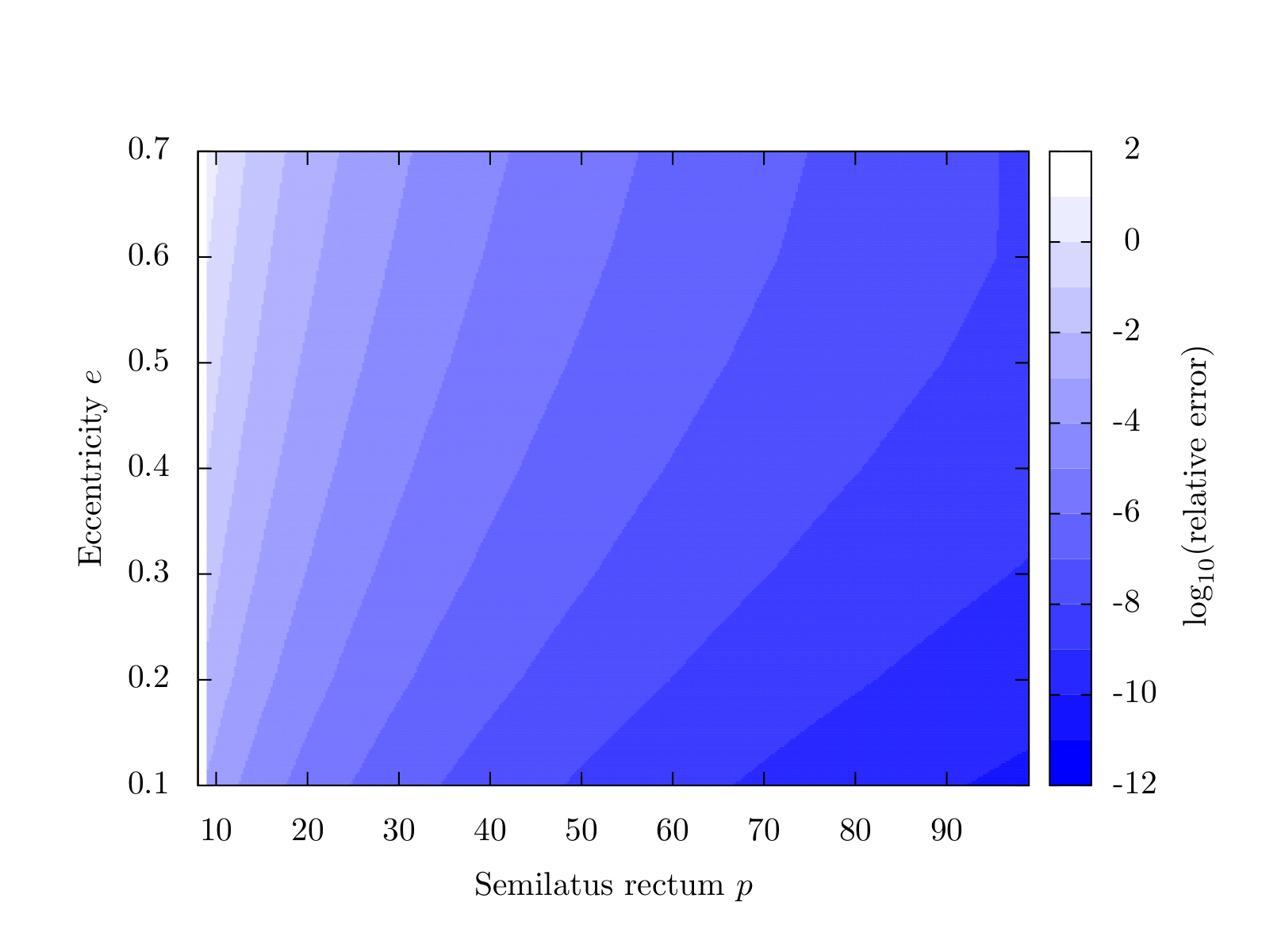}
\caption{Strong-field comparison between the 7PN expansion and energy fluxes 
computed with a Lorenz gauge/RWZ gauge hybrid self-force code \cite{OsbuETC14} 
(courtesy T.~Osburn).
\label{fig:strongField}}
\end{figure}

The 7PN fit was obtained using orbits with eccentricities between $0.0$ 
and $0.1$, and using orbital separations of $p = 10^{10}$ through 
$p = 10^{35}$.  A natural question to ask is how well does the PN expansion 
work if we compute fluxes from higher eccentricity orbits and from orbits 
with much smaller separations?  The answer is: quite well.  
Fig.~\ref{fig:twoPanelPlot} shows (on the left) the circular orbit limit normalized 
energy flux (dominated by the Peters-Mathews term) as black points, and 
the red curve is the fit from our 7PN model.  Here we have reduced the 
orbital separation to $p = 10^3$ and we compare the data and model all the 
way up to $e = 0.6$.  On the right side we show the effect of subtracting 
the model containing all terms up to and including the 6PN contributions.  
With an orbit with a radius of $p = 10^3$, the residuals have dropped by 
14 orders of magnitude.  The remaining part of the model (6.5PN and 7PN) is 
then shown to still fit these residuals.

We examined the fit then at still smaller radius orbits.  Figure 
\ref{fig:strongField} compares our 7PN determination to energy fluxes 
obtained by T.~Osburn using a Lorenz gauge/RWZ gauge hybrid code 
\cite{OsbuETC14}.  Energy fluxes have accuracies of $10^{-3}$ all the way in 
as close as $p = 30$.

\section{Conclusions}
\label{sec:conclusions}

In this paper we have presented a first set of results from a new 
eccentric-orbit MST code.  The code, written in \emph{Mathematica}, combines 
the MST formalism and arbitrary-precision functions to solve the perturbation 
equations to an accuracy of 200 digits.  We computed the energy flux at 
infinity, at lowest order in the mass ratio (i.e., in the region of parameter 
space overlapped by BHP and PN theories).  In this effort, we computed 
results for approximately 1,700 distinct orbits, with up to as many as 7,400 
Fourier-harmonic modes per orbit.  

The project had several principal 
new results.  First, we confirmed previously computed PN flux expressions 
through 3PN order.  Second, in the process of this analysis, we developed 
a procedure and new high-order series expansions for the non-closed form 
hereditary terms at 1.5PN, 2.5PN, and 3PN order.  Significantly, at 2.5PN 
order we overcame some of the previous roadblocks to writing down accurate 
high-order expansions for this flux contribution (App.~\ref{sec:massQuad}).  
The 3PN hereditary term was shown to have a subtle singular behavior as 
$e_t \to 1$.  All of this clarification of the behavior of the hereditary 
terms was aided by an asymptotic analysis of a set of enhancement functions.  
In the process we were able to predict the form of eccentricity singular 
factors that appear at each PN order.  Third, based on that understanding, 
we then used the high accuracy of the code to find a mixture of new analytic 
and numeric flux terms between 3.5PN and 7PN.  We built in expected forms 
for the eccentricity singular factors, allowing the determined power series 
in $e^2$ to better determine the flux at high values of $e$.  

The code we have developed for this project can be used not only to compute 
fluxes but also local GSF quantities.  Recently Akcay et al.~\cite{AkcaETC15} 
made comparisons between GSF and PN values of the eccentric orbit 
generalization of Detweiler's redshift invariant 
($\Delta U$) \cite{Detw08,BaraSago11}.  We may be able to extend these 
comparisons beyond the current 4PN level and compute currently unknown 
coefficients (in the linear in $\mu/M$ limit).  We can also modify the code 
to make calculations on a Kerr background.

\acknowledgments

The authors thank T.~Osburn, A.~Shah, B.~Whiting, S.A.~Hughes, 
N.K.~Johnson-McDaniel, and L.~Barack for helpful discussions.  We also 
thank L.~Blanchet and an anonymous referee for separately asking several 
questions that led us to include a section on asymptotic analysis of 
enhancement functions and eccentricity singular factors.  This work was 
supported in part by NSF grant PHY-1506182.  EF acknowledges support from 
the Royster Society of Fellows at the University of North Carolina-Chapel 
Hill.  CRE is grateful for the hospitality of the Kavli Institute for 
Theoretical Physics at UCSB (which is supported in part by the National 
Science Foundation under Grant No. NSF PHY11-25915) and the Albert Einstein 
Institute in Golm, Germany, where part of this work was initiated.  CRE also 
acknowledges support from the Bahnson Fund at the University of North 
Carolina-Chapel Hill.  SH acknowledges support from the Albert Einstein 
Institute and also from Science Foundation Ireland under Grant 
No.~10/RFP/PHY2847.  SH also acknowledges financial support provided under 
the European Union's H2020 ERC Consolidator Grant ``Matter and strong-field 
gravity: New frontiers in Einstein’s theory'' grant agreement 
no.~MaGRaTh--646597.

\appendix

\section{The mass quadrupole tail at 1PN}
\label{sec:massQuad}

In their Eqn.~(5.14) Arun et al.~\cite{ArunETC08b} write the tail contribution
to the mass quadrupole flux in terms of enhancement factors as
\begin{align}
\begin{split}
&\mathcal{F}_\text{tail}^\text{mass quad} = 
\frac{32}{5} \nu^2 y^{13/2}	\\
& \hspace{2ex}
\times \left\{
4 \pi \vp (e_t) 
+ \pi y \left[ -\frac{428}{21} \a (e_t) + \frac{178}{21} \nu \th (e_t) \right]
\right\},
\end{split}
\end{align}
where $\nu$ is the symmetric mass ratio. The $\alpha$ and $\th$ enhancement
factors contribute at 2.5PN and
depend on the 1PN mass quadrupole. This dependence makes them more difficult to
calculate than any of the other enhancement factors up to 3PN.
Arun et al.~outline a procedure for computing $\alpha$ and $\theta$ numerically,
showing their results graphically.

In this appendix we summarize our calculation of the 2.5PN enhancement
factor $\alpha$, which contributes to $\psi$ 
[see Arun et al.~Eqn.~(6.1a)]. Our presentation follows closely that given
in Sec.~IVD of Arun et al. Unlike them, we work in the $\mu/M\to 0$
limit and use the BHP notation already established in this paper. 
Significantly, we were able to find an analytic expression for $\alpha$ 
as a high-order power-series in eccentricity.  We give this series (with a 
singular factor removed) to 20th-order in Sec.~\ref{sec:preparePN}, but we 
have computed it to 70th order.  Although we are working in the 
$\mu/M\to 0$ limit, it may be possible to employ the method outlined here 
to obtain the finite mass-ratio term $\th$ in a similar power series.

\subsection{Details of the flux calculation}

With $y$ expanded in $p$ as discussed in the text, we are able to find
1PN expansions of $\mathcal{E}$ and $\mathcal{L}$ using 
Eqn.~\eqref{eqn:defeandp}.  In order to find the other orbital quantities 
that will go into the 1PN mass quadrupole we use the quasi-Keplerian (QK) 
parametrization \cite{DamoDeru85}.  As that parametrization is well covered 
in Ref.~\cite{ArunETC08b} and many other papers we will not go into detail 
here, except to make two points. 

First, in the QK parametrization $r_p$, $\vp_p$ and their derivatives are
expressed as functions of the eccentric anomaly $u$.  As such, when computing 
the Fourier series coefficients of the mass quadrupole we perform integrations
with respect to $u$.  Additionally, we note that when using the QK 
parametrization, at 1PN there are 3 eccentricities $e_t$, $e_r$ and $e_\vp$.
Typically eccentric orbits are described using $e_t$.  Through 3PN $e$ and 
$e_t$ are related to each other via \eqref{eqn:etToe}.  We use this 
expression to convert known PN enhancement factors to $e$ dependence, as 
shown in Sec.~\ref{sec:EnhanceDarwin}.

To lowest order in $\mu/M$, the 1PN mass quadrupole is given by
\begin{align}
\begin{split}
I_{ij} &= \mu \bigg[
\l 1 + \frac{29}{42} \dot{x}_p^2 
- \frac{5}{7} \frac{M}{r_p}  \r x_p^{\langle i} x_p^{j\rangle}
 \\
& \hspace{14ex} + \frac{11}{21} r_p^2 \dot{x}_p^{\langle i} 
\dot{x}_p^{j\rangle} 
- \frac{4}{7} r_p \dot{r}_p x_p^{\langle i} \dot{x}_p^{j\rangle}
\bigg],
\end{split}
\end{align}
with the angle brackets indicating a symmetric trace free projection 
(e.g. $x_p^{\langle i} x_p^{j\rangle} = x_p^{i} x_p^{j}-r_p^2 \d^{ij} /3 $).
We start by expanding $I_{ij}$ in a Fourier series 
\be
\label{eqn:Iij1PN}
I_{ij}(t) = \sum_{n=-\infty}^\infty \sum_{m=-2}^2
\underset{(m,n)\ }{\mathcal{I}_{ij}}\ e^{i \o_{mn} t}.
\ee
Note that in this series we are following the sign convention of 
Arun et al.~which differs from that in, e.g. Eqn.~\eqref{eqn:psiSeries}.
Arun et al. give the mass quadrupole tail flux in their Eqn.~(4.17) as
\begin{align}
\label{eqn:MassQuad1PN}
&\mathcal{F}_\text{tail}^\text{mass quad} = \\
& \frac{4M}{5}
\left\langle I_{ij}^{(3)}
(t)\int_0^\infty d\tau I_{ij}^{(5)}(t-\tau)\left[\log\left(\frac{\tau}{2r_0}\right)
+\frac{11}{12}\right]\right\rangle. \notag 
\end{align}
Here the superscripts (3) and (5) indicate the number of time derivatives,
and $r_0$ is a constant with dimensions length which does not appear in the 
final expression for the flux.  Inserting Eqn.~\eqref{eqn:Iij1PN} we find
\begin{align}
\label{eqn:massQuadSumFull}
& \mathcal{F}_\text{tail}^\text{mass quad} = 
- \frac{4M}{5} \sum_{m,m',n,n'} \o_{mn}^3\o_{m'n'}^5  \\
& \hspace{8ex} 
\times \underset{(m,n)\ }{\mathcal{I}_{ij}} 
\underset{(m',n')\ }{\mathcal{I}^*_{ij}}
\left\langle e^{i\left[ \O_\vp(m-m') + \O_r (n-n')\right]t} 
\right\rangle \notag
\\
&\hspace{8ex} \times
\int_0^\infty d\tau \ e^{i\o_{m'n'}\tau}
\left[\log\left(\frac{\tau}{2r_0}\right)+\frac{11}{12}\right]. \notag
\end{align}
In deriving Eqn.~\eqref{eqn:massQuadSumFull} we have reversed the sign on 
both $m'$ and $n'$ and used the crossing relation 
$_{(m,n)}\mathcal{I}_{ij}= {_{(-m,-n)}\mathcal{I}^*_{ij}}$.
We split this expression up into four terms, writing
\be
\label{eqn:massQuadSum}
\mathcal{F}_\text{tail}^\text{mass quad} =
\sum_{n,n',m,m'}\mathcal{A}\cdot\mathcal{B}\cdot
\mathcal{C}\cdot\mathcal{D},
\ee
with
\begin{align}
\begin{split}
\mathcal{A} &=  -\frac{4M}{5} \o_{mn}^3\o_{m'n'}^5 ,\\
\mathcal{B} &= 
\underset{(m,n)\ }{\mathcal{I}_{ij}} \underset{(m',n')\ }{\mathcal{I}^*_{ij}},
\\
\mathcal{C} &= 
\left\langle e^{i\left[ \O_\vp(m-m') + \O_r (n-n')\right]t} \right\rangle,\\
\mathcal{D} &= \int_0^\infty d\tau \ e^{i\o_{m'n'}\tau}
\left[\log\left(\frac{\tau}{2r_0}\right)+\frac{11}{12}\right].
\end{split}
\end{align}
Each of these terms has 0PN and 1PN contributions. For example,
$\mathcal{A} = \mathcal{A}_0 + y \mathcal{A}_1$, and similarly for 
$\mathcal{B}$, $\mathcal{C}$, and $\mathcal{D}$.
Then, through 1PN the summand in Eqn.~\eqref{eqn:massQuadSum} is
\begin{align}
\label{eqn:massQuadSummand}
\mathcal{A}\cdot\mathcal{B}\cdot\mathcal{C}\cdot\mathcal{D}
&= \mathcal{A}_0\mathcal{B}_0\mathcal{C}_0\mathcal{D}_0
+ y
\bigg(\mathcal{A}_1\mathcal{B}_0\mathcal{C}_0\mathcal{D}_0 \\
&
+ \mathcal{A}_0\mathcal{B}_1\mathcal{C}_0\mathcal{D}_0 
+ \mathcal{A}_0\mathcal{B}_0\mathcal{C}_1\mathcal{D}_0 
+ \mathcal{A}_0\mathcal{B}_0\mathcal{C}_0\mathcal{D}_1 \bigg). \notag
\end{align}
Expanding $\mathcal{A}$ we find
\begin{align}
\begin{split}
\mathcal{A}_0 &= -\frac{4 y^{12}}{5M^7} n^3n'^5, \\
\mathcal{A}_1 &= -\frac{12 y^{12}}{5M^7} 
\frac{ \l 3 m n^2 n'^5 + 5 m' n^3 n'^4 - 8 n^3 n'^5 \r}{1-e_t^2}
\end{split}
\end{align}
We next consider the moments in $\mathcal{B}$, writing
\be
\underset{(m,n)\ }{\mathcal{I}_{ij}} = \underset{(m,n)\ }{\mathcal{I}_{ij}^{0}}
+y \underset{(m,n)\ }{\mathcal{I}_{ij}^{1}},
\ee
and hence
\begin{align}
\begin{split}
\mathcal{B}_0 &= \underset{(m,n)\ }{\mathcal{I}_{ij}^{0}}
\underset{(m',n')\ }{\mathcal{I}_{ij}^{0*}}, \\
\mathcal{B}_1 &= \underset{(m,n)\ }{\mathcal{I}_{ij}^{0}}
\underset{(m',n')\ }{\mathcal{I}_{ij}^{1*}} 
+ \underset{(m,n)\ }{\mathcal{I}_{ij}^{1}}
\underset{(m',n')\ }{\mathcal{I}_{ij}^{0*}}.
\label{eqn:B1Expand}
\end{split}
\end{align}
The heart of the calculation of $\alpha$ comes down to computing the 
Fourier coefficients $_{(m,n)}{\mathcal{I}_{ij}^{1}}$ in 
Eqn.~\eqref{eqn:B1Expand}. As mentioned above, we compute these
terms by representing the elements of $I_{ij}$ in the QK parametrization.
The Fourier coefficients are then computed by integrating
with respect to the eccentric anomaly $u$. While we 
cannot perform these integrals for completely generic expressions, we do find
that we can expand $I_{ij}$ in eccentricity 
and obtain $_{(m,n)}{\mathcal{I}_{ij}^{1}}$
as a power series in $e_t$. Furthermore, as shown in Sec.~\ref{sec:preparePN}
we are able to remove singular factors in this expansion, leading to much 
improved convergence for large eccentricity.  Significantly, we find that 
$\mathcal{B}_0$ and $\mathcal{B}_1$ are only nonzero when $m=m'$.

Next we consider $\mathcal{C}$. Expanding the complex
exponential to 1PN, we can perform the time-average integral
and we find
\begin{align}
\begin{split}	
\label{eqn:Cterms}
\mathcal{C}_0 &=\delta_{n,n'}, \\
\mathcal{C}_1 &= i\pi(m-m')\delta_{n,n'}+\frac{m-m'}{n-n'}\left(
1-\delta_{n,n'}\right),
\end{split}
\end{align}
where the $1-\d_{n,n'}$ indicates that the second term vanishes when $n=n'$.
The case where we employ $\mathcal{C}_0 = \delta_{n,n'}$
greatly simplifies the calculation, taking us from a doubly-infinite
sum to a singly-infinite sum. Remarkably, the $\mathcal{C}_1$ term
does not contribute at all. This follows from the fact that 
it is proportional to $m-m'$ and the $B_0$ terms only contribute 
when $m=m'$. Thus, the doubly-infinite sum found by Arun et al.~reduces
to a singly-infinite sum (at least) in the limit that $\mu/M \to 0$ at 1PN.

The tail integral for $\mathcal{D}$ is computed using expressions in 
Ref.~\cite{ArunETC08b}. Each of the terms $\mathcal{D}_0$ and
$\mathcal{D}_1$ 
is complex and we find that the imaginary part cancels
after summing over positive and negative $m'$ and $n'$, leaving a 
purely real contribution to the flux. The real contributions to these terms are
\begin{align}
\mathcal{D}_0 =
- \frac{M}{y^{3/2}} \frac{\pi}{2|n'|}, \quad
\mathcal{D}_1 = 
 \frac{3 \pi M (n' - m')}{2 y^{3/2} (1-e_t^2) n' |n'|}.
\end{align}

At this point we combine the 0PN and 1PN contributions to 
$\mathcal{A}$, $\mathcal{B}$, $\mathcal{C}$, and $\mathcal{D}$ in 
Eqns.~\eqref{eqn:massQuadSummand} and \eqref{eqn:massQuadSum}.
The Kronecker deltas in Eqn.~\eqref{eqn:Cterms} along with the
fact that $\mathcal{B}$ is nonzero only for $m=m'$ reduces the sum to
\be
\mathcal{F}_\text{tail}^\text{mass quad} =
\sum_{n=-\infty}^\infty \sum_{m=-2}^2
\mathcal{A}\cdot\mathcal{B}\cdot
\mathcal{C}\cdot\mathcal{D}.
\ee
Furthermore, expanding the Fourier coefficients $_{(m,n)}{\mathcal{I}_{ij}^{1}}$
in eccentricity to some finite order truncates the sum over $n$.
This sum yields both the 1.5PN enhancement factor $\vp$ and the 2.5PN 
factor $\alpha$.

\section{Solving for $\nu$}
\label{sec:solveNu}

The full forms of the $\alpha_n^\nu$, $\beta_n^\nu$, and $\gamma_n^\nu$ 
mentioned in \ref{sec:innerMST} are

\begin{align}
\alpha_n^\nu &= 
\frac{i\epsilon\kappa(n+\nu+1+s+i\epsilon)}{(n+\nu+1)(2n+2\nu+3)}\notag
\\&\times(n+\nu+1+s-i\epsilon)
(n+\nu+1+i\tau),\notag\\
\beta_n^\nu &=-\lambda-s(s+1)+(n+\nu)(n+\nu+1)+\epsilon^2+
\epsilon(\epsilon-mq)\notag\\&+\frac{\epsilon(\epsilon-mq)(s^2+\epsilon^2)}
{(n+\nu)(n+\nu+1)},\notag\\
\gamma_n^\nu&=-\frac{i\epsilon\kappa(n+\nu-s+i\epsilon)
(n+\nu-s-i\epsilon)(n+\nu-i\tau)}{(n+\nu)(2n+2\nu-1)}.
\end{align}
Now introduce the continued fractions
\begin{align}
\label{eqn:RL}
R_n^\nu\equiv\frac{a_n^\nu}{a_{n-1}^\nu}=-\frac{\gamma_n^\nu}{\beta_n^\nu-}
\frac{\alpha_n^\nu\gamma_{n+1}^\nu}{\beta_{n+1}^\nu-}\frac{\alpha_{n+1}^\nu
\gamma_{n+2}^\nu}{\beta_{n+2}^\nu-}\cdots,\notag\\
L_n^\nu\equiv\frac{a_n^\nu}{a_{n+1}^\nu}=-\frac{\alpha_n^\nu}{\beta_n^\nu-}
\frac{\alpha_{n-1}^\nu\gamma_n^\nu}{\beta_{n-1}^\nu-}\frac{\alpha_{n-2}^\nu
\gamma_{n-1}^\nu}{\beta_{n-2}^\nu-}\cdots.
\end{align}
Then recall that the series coefficients $\{a_n^\nu\}$ satisfy the
 three-term
recurrence \eqref{eqn:recurrence}, which can now be rewritten as
\be
\beta_n^\nu+\alpha_n^\nu R_{n+1}+\gamma_n^\nu L_{n-1}=0.
\ee
This holds for arbitrary $n$, so in particular we can set $n=0$,
\be
\label{eqn:nuRoot}
\beta_0^\nu+\alpha_0^\nu R_{1}+\gamma_0^\nu L_{-1}=0.
\ee
In practice, $\nu$ is determined by numerically looking 
for the roots of 
\eqref{eqn:nuRoot}. Formally, $R_n^\nu$ and $L_n^\nu$ 
have an infinite depth, 
but may be truncated at finite depth in \eqref{eqn:nuRoot}
depending on the 
precision to which it is necessary to determine $\nu$. 

We note also that there exists a low-frequency
expansion for $\nu$. Letting $\epsilon=2M\omega$,
\begin{align*}
\nu=&l+\frac{1}{2l+1}\bigg(
-2-\frac{s^2}{l(l+1)}+\frac{\left[(l+1)^2-s^2\right]^2}{(2l+1)(2l+2)(2l+3)}
\notag\\&-\frac{(l^2-s^2)^2}{(2l-1)(2l)(2l+1)}
\bigg)\epsilon^2
+\mathcal{O}(\epsilon^4).
\end{align*}
For given $l$, we are able to take this expansion
to arbitrary order, and therefore easily and
quickly determine $\nu$ to very high precision
for small frequencies. 

\section{Analytic and numeric coefficients in the high-order post-Newtonian 
functions}
\label{sec:numericEnh}

Numerical values for the remaining coefficients in the high-order PN functions 
\eqref{eqn:Ienh72}-\eqref{eqn:Ienh7L2} are provided in Tables 
\ref{tab:35Decimal}-\ref{tab:7L2Decimal}.

\begin{center}
\begin{table}[h]
\caption{Coefficients in the 3.5PN function according to the form of
Eqn.~\eqref{eqn:Ienh72} for
which we were unable to determine rational forms.  
\label{tab:35Decimal}} 
\begin{tabular}{c|c}
\hline
\hline
\multicolumn{2}{ c }{3.5PN Coefficients}\\
\hline
Coefficient & Decimal Form \\ \hline
$b_{26}$ & $-1.490561574783877$
\\\hline
$b_{28}$ & $-1.193065651880562$
\\\hline
$b_{30}$ & $-0.9727274285773439$
\\\hline
$b_{32}$ & $-0.8055449753687710$
\\\hline
$b_{34}$ & $-0.6760669725730609$
\\\hline
$b_{36}$ & $-0.5740065189983486$
\\\hline
$b_{38}$ & $-0.4923153284656641$
\\\hline
$b_{40}$ & $-0.4260425626411813$
\\\hline
$b_{42}$ & $-0.3716342707794599$
\\\hline
$b_{44}$ & $-0.3264898836852403$
\\\hline
$b_{46}$ & $-0.2886737566317485$
\\\hline
$b_{48}$ & $-0.2567230158706580$
\\\hline
$b_{50}$ & $-0.229516796538214$
\\\hline
$b_{52}$ & $-0.2061855361551$
\\\hline
$b_{54}$ & $-0.1860469549$
\\\hline
\hline
\end{tabular}
\end{table}
\end{center}

\begin{center}
\begin{table}[h]
\caption{Coefficients in the 4PN non-log function $\mathcal{L}_{4}$ according 
to the form of Eqn.~\eqref{eqn:Ienh4}. 
\label{tab:4NLDecimal}} 
\begin{tabular}{c|c}
\hline
\hline
\multicolumn{2}{ c }{4PN Non-Log Coefficients}\\
\hline
Coefficient & Decimal Form \\
\hline
$d_{8}$ & -12385.51003537713
\\\hline
$d_{10}$ & 5863.111480566811
\\\hline
$d_{12}$ & 3622.327433443339
\\\hline
$d_{14}$ & 2553.863176036157
\\\hline
$d_{16}$ & 2026.951300184891
\\\hline
$d_{18}$ & 1688.454045610002
\\\hline
$d_{20}$ & 1449.705886053665
\\\hline
$d_{22}$ & 1271.358072870572
\\\hline
$d_{24}$ & 1132.705339539895
\\\hline
$d_{26}$ & 1021.659868559411
\\\hline
$d_{28}$ & 930.6413570026334
\\\hline
$d_{30}$ & 854.6360714818981
\\\hline
$d_{32}$ & 790.1870990544641
\\\hline
$d_{34}$ & 734.8297131797
\\\hline
$d_{36}$ & 686.7572696
\\\hline
$d_{38}$ & 644.61426
\\\hline
$d_{40}$ & 607.363
\\\hline
\hline
\end{tabular}
\end{table}
\end{center}

\begin{center}
\begin{table}[h]
\caption{Coefficients in the 4.5PN non-log function $\mathcal{L}_{9/2}$ according to the
form of Eqn.~\eqref{eqn:Ienh92}. 
\label{tab:45NLDecimal}} 
\begin{tabular}{c|c}
\hline
\hline
\multicolumn{2}{ c }{4.5PN Non-Log Coefficients}\\
\hline
Coefficient & Decimal Form \\
\hline
$h_{4}$ & $34992.49298556799$
\\\hline
$h_{6}$ & $39847.21900599596$
\\\hline
$h_{8}$ & $-6749.004994216654$
\\\hline
$h_{10}$ & $-11556.5757682355$
\\\hline
$h_{12}$ & $-3380.99341129105$
\\\hline
$h_{14}$ & $-2106.48821413306$
\\\hline
$h_{16}$ & $-1605.13441975986$
\\\hline
$h_{18}$ & $-1305.76892323295$
\\\hline
$h_{20}$ & $-1104.20944659088$
\\\hline
$h_{22}$ & $-958.287100012754$
\\\hline
$h_{24}$ & $-847.360081662457$
\\\hline
$h_{26}$ & $-759.987768729771$
\\\hline
$h_{28}$ & $-689.280637106089$
\\\hline
$h_{30}$ & $-630.824342070861$
\\\hline
$h_{32}$ & $-581.651645011457$
\\\hline
$h_{34}$ & $-539.69025201625$
\\\hline
$h_{36}$ & $-503.44686180$
\\\hline
\hline
\end{tabular}
\end{table}
\end{center}

\begin{center}
\begin{table}[h]
\caption{Coefficients in the 4.5PN log function 
$\mathcal{L}_{9/2L}$ according to the form of
Eqn.~\eqref{eqn:Ienh92L}.
\label{tab:45LDecimal}}
\begin{tabular}{c|c}
\hline\hline
\multicolumn{2}{ c }{4.5PN Log Coefficients}\\
\hline
Coefficient & Decimal Form \\
\hline
$g_{20}$ & $0.00005237112441246353$
\\\hline
$g_{22}$ & $1.500221637093169\times 10^{-6}$
\\\hline
$g_{24}$ & $-1.445846088397795\times 10^{-6}$
\\\hline
$g_{26}$ & $-3.761668215289038\times 10^{-7}$
\\\hline
$g_{28}$ & $-6.09085540489931\times 10^{-8}$
\\\hline
$g_{30}$ & $-8.1161948669\times 10^{-9}$\\
\hline
\hline
\end{tabular}
\end{table}
\end{center}

\begin{center}
\begin{table}[h]
\caption{Coefficients in the 5PN non-log function $\mathcal{L}_{5}$ according to the form of
equation Eqn.~\eqref{eqn:Ienh5}. 
\label{tab:5NLDecimal}} 
\begin{tabular}{c|c}
\hline
\hline
\multicolumn{2}{ c }{5PN Non-Log Coefficients}\\
\hline
Coefficient & Decimal Form \\ \hline
$k_{2}$ & $-30239.82434287415$
\\\hline
$k_{4}$ & $-104593.9352475713$
\\\hline
$k_{6}$ & $75463.00111171990$
\\\hline
$k_{8}$ & $269065.9776584819$
\\\hline
$k_{10}$ & $73605.36477482046$
\\\hline
$k_{12}$ & $-41367.54579072399$
\\\hline
$k_{14}$ & $-30242.01469796259$
\\\hline
$k_{16}$ & $-22936.05821746133$
\\\hline
$k_{18}$ & $-18706.03853031426$
\\\hline
$k_{20}$ & $-15829.44370858490$
\\\hline
$k_{22}$ & $-13730.52183792896$
\\\hline
$k_{24}$ & $-12127.0027508276$
\\\hline
$k_{26}$ & $-10860.422505$
\\\hline
\hline
\end{tabular}
\end{table}
\end{center}

\begin{center}
\begin{table}[h]
\caption{Coefficients in the 5PN log function
$\mathcal{L}_{5L}$ according to the
form of Eqn.~\eqref{eqn:Ienh5L}. 
\label{tab:5LDecimal}} 
\begin{tabular}{c|c}
\hline
\hline
\multicolumn{2}{ c }{5PN Log Coefficients}\\
\hline
Coefficient & Decimal Form \\ \hline
$j_{26}$ & $-17.11762038204405$
\\\hline
$j_{28}$ & $-13.5361663264$
\\\hline
\hline
\end{tabular}
\end{table}
\end{center}

\begin{center}
\begin{table}[h]
\caption{Coefficients in the 5.5PN non-log function $\mathcal{L}_{11/2}$ according to the
form of Eqn.~\eqref{eqn:Ienh112}. 
\label{tab:55NLDecimal}} 
\begin{tabular}{c|c}
\hline
\hline
\multicolumn{2}{ c }{5.5PN Non-Log Coefficients}\\
\hline
Coefficient & Decimal Form \\ \hline
$m_{2}$ & $-6745.196131429910$
\\\hline
$m_{4}$ & $-141316.5392653812$
\\\hline
$m_{6}$ & $-469215.6608702663$
\\\hline
$m_{8}$ & $-434086.0636932912$
\\\hline
$m_{10}$ & $48474.32085716919$
\\\hline
$m_{12}$ & $129328.2556268356$
\\\hline
$m_{14}$ & $72095.03819665170$
\\\hline
$m_{16}$ & $53240.504083351$
\\\hline
$m_{18}$ & $42978.60994$
\\\hline
\hline
\end{tabular}
\end{table}
\end{center}

\begin{center}
\begin{table}[h]
\caption{Coefficients in the 5.5PN log function
$\mathcal{L}_{11/2L}$ according to the form of
Eqn.~\eqref{eqn:Ienh112L}. 
\label{tab:55LDecimal}} 
\begin{tabular}{c|c}
\hline
\hline
\multicolumn{2}{ c }{5.5PN Log Coefficients}\\
\hline
Coefficient & Decimal Form \\ \hline
$l_{12}$ & $189.3698825402159$
\\\hline
$l_{14}$ & $0.007977195397456623$
\\\hline
$l_{16}$ & $0.0239041225377$
\\\hline
\hline
\end{tabular}
\end{table}
\end{center}

\begin{center}
\begin{table}[h]
\caption{Coefficients in the 6PN non-log function $\mathcal{L}_{6}$ according to the form of
Eqn.~\eqref{eqn:Ienh6}. 
\label{tab:6PNDecimal}} 
\begin{tabular}{c|c}
\hline
\hline
\multicolumn{2}{ c }{6PN Non-Log Coefficients}\\
\hline
Coefficient & Decimal Form \\ \hline
$n_{2}$ & $183747.5626219038$
\\\hline
$n_{4}$ & $1960389.869256877$
\\\hline
$n_{6}$ & $5197796.234905070$
\\\hline
$n_{8}$ & $2705571.654688992$
\\\hline
$n_{10}$ & $-1438997.974016927$
\\\hline
$n_{12}$ & $-923762.260160$
\\\hline
\hline
\end{tabular}
\end{table}
\end{center}

\begin{center}
\begin{table}[h]
\caption{Coefficients in the 6PN log function $\mathcal{L}_{6L}$ according to the form of
Eqn.~\eqref{eqn:Ienh6L}. 
\label{tab:6LDecimal}} 
\begin{tabular}{c|c}
\hline
\hline
\multicolumn{2}{ c }{6PN Log Coefficients}\\
\hline
Coefficient & Decimal Form \\ \hline
$p_2$ & $-45332.65954460482$
\\\hline
$p_4$ & $-339065.8391750689$
\\\hline
$p_6$ & $-612772.4442656394$
\\\hline
$p_8$ & $-63054.18877810686$
\\\hline
$p_{10}$ & $298247.27730847$
\\\hline
$p_{12}$ & $107300.427923$
\\\hline
\end{tabular}
\end{table}
\end{center}

\begin{center}
\begin{table}[h]
\caption{Coefficients in the 6.5PN non-log function $\mathcal{L}_{13/2}$ according to the form of
Eqn.~\eqref{eqn:Ienh132}. 
\label{tab:65NLDecimal}} 
\begin{tabular}{c|c}
\hline
\hline
\multicolumn{2}{ c }{6.5PN Non-Log Coefficients}\\
\hline
Coefficient & Decimal Form \\ \hline
$r_2$ & $-127378.8277728414$
\\\hline
$r_4$ & $-1087743.1117750650$
\\\hline
$r_6$ & $-1820396.3219686398$
\\\hline
$r_8$ & $837531.195260$
\\\hline
$r_{10}$ & $2191527.42$
\\\hline
$r_{12}$ & $-676000$
\\\hline
\end{tabular}
\end{table}
\end{center}

\begin{center}
\begin{table}[h]
\caption{Coefficients in the 6.5PN log function $\mathcal{L}_{13/2L}$ according to the form of
Eqn.~\eqref{eqn:Ienh132L}. 
\label{tab:65LDecimal}} 
\begin{tabular}{c|c}
\hline
\hline
\multicolumn{2}{ c }{6.5PN Log Coefficients}\\
\hline
Coefficient & Decimal Form \\ \hline
$s_2$ & $3556.505482510765$
\\\hline
$s_4$ & $-88650.24536994091$
\\\hline
$s_6$ & $-725275.3694386568$
\\\hline
$s_8$ & $-1383744.78113972$
\\\hline
$s_{10}$ & $-775068.8299$
\\\hline
$s_{12}$ & $-117673$
\\\hline
\end{tabular}
\end{table}
\end{center}

\begin{center}
\begin{table}[h]
\caption{Coefficients in the 7PN non-log function $\mathcal{L}_7$ according to the form of
Eqn.~\eqref{eqn:Ienh7}.
\label{tab:7NLDecimal}} 
\begin{tabular}{c|c}
\hline
\hline
\multicolumn{2}{ c }{7PN Non-Log Coefficients}\\
\hline
Coefficient & Decimal Form \\ \hline
$t_2$ & $274599.5315289643$
\\\hline
$t_4$ & $-1830774.789990$
\\\hline
$t_6$ & $-22622781.31$
\\\hline
\end{tabular}
\end{table}
\end{center}

\begin{center}
\begin{table}[h]
\caption{Coefficients in the 7PN log function $\mathcal{L}_{7L}$ according to the form of
Eqn.~\eqref{eqn:Ienh7L}.
\label{tab:7LDecimal}} 
\begin{tabular}{c|c}
\hline
\hline
\multicolumn{2}{ c }{7PN Log Coefficients}\\
\hline
Coefficient & Decimal Form \\ \hline
$u_2$ & $196177.8173496888$
\\\hline
$u_4$ & $3290034.8114615$
\\\hline
$u_6$ & $13274887.225$
\\\hline
$u_8$ & $17171440$
\\\hline
\end{tabular}
\end{table}
\end{center}

\begin{center}
\begin{table}[h]
\caption{Coefficients in the 7PN log-squared function $\mathcal{L}_{7L^2}$ according to the form of
Eqn.~\eqref{eqn:Ienh7L2}. 
\label{tab:7L2Decimal}} 
\begin{tabular}{c|c}
\hline
\hline
\multicolumn{2}{ c }{7PN Log-Squared Coefficients}\\
\hline
Coefficient & Decimal Form \\ \hline
$v_2$ & $11387.49469684699$
\\\hline
$v_4$ & $157165.280283919$
\\\hline
$v_6$ & $568228.86246$
\\\hline
$v_8$ & $677671.2$
\\\hline
\end{tabular}
\end{table}
\end{center}

\clearpage

\bibliography{mstflux}

\end{document}